\documentclass[journal=cmatex,manuscript=article,longbibliography]{achemso}
\usepackage{chemformula} % Formula subscripts using \ch{}
\usepackage[T1]{fontenc} % Use modern font encodings

\usepackage{amsfonts,mathrsfs,amsmath,amsthm,amssymb,bbold}
\usepackage{hyperref}
\usepackage{xcolor}                  % for \textcolor
\usepackage[normalem]{ulem}     
\usepackage{epsfig}
\usepackage{bm,mathrsfs}
\usepackage{lineno,hyperref}
\usepackage{booktabs}
\usepackage{booktabs,tabularx,makecell} 
\usepackage{pdfpages}
\newcommand{\mobunit}{\ensuremath{\mathrm{cm}^2\,\mathrm{V}^{-1}\,\mathrm{s}^{-1}}}
\newcolumntype{C}[1]{>{\centering\arraybackslash}p{#1}}

\newcommand*{\si}{\textrm{the Supporting Information}}
\newcommand{\etal}{\textit{et al.~}}

\author{Thi Ngoc Huyen Vu}
\affiliation{Institute for Materials Research, Tohoku University, Sendai, Miyagi 980--8577, Japan}

\author{Yu Kumagai}
\affiliation{Institute for Materials Research, Tohoku University, Sendai, Miyagi 980--8577, Japan}
\alsoaffiliation{Division for the Establishment of Frontier Sciences, Organization for Advanced Studies, Tohoku University, Sendai, Miyagi 980--8577, Japan}
\email{yukumagai@tohoku.ac.jp}
\title{Investigation of Hole Dopability in Oxygen $2p$-Dominated Bands}

\begin{document}  

%\begin{tocentry}
%\includegraphics[width=8.45cm]{TOC.eps}
%\end{tocentry}

\begin{abstract}
    The development of $p$-type oxide semiconductors remains impeded
    by the inherently low-lying valence-band maximum (VBM) dominated by O-2$p$ states.
    A prevailing approach to mitigate this limitation is
    to elevate the VBM by introducing cation states that hybridize with O-2$p$ orbitals
    or lie energetically above the O-2$p$ level.
    Nevertheless, the $p$-type oxides reported to date exhibit limited hole mobilities.
    To expand the search space, it is essential to accurately understand the intrinsic difficulty
    of introducing holes into O-2$p$-dominated bands.
    Accordingly, we evaluated 845 oxides to identify those
    in which holes can be doped into O-2$p$-dominated bands.
    Our high-throughput screening revealed CaCdO$_2$ as the only promising exemplar,
    in which the VBM is slightly hybridized with deep-lying Cd-3$d$ states.
    Our screening suggests that hole doping into O-2$p$-dominated bands is extremely difficult
    and thus reinforces the effectiveness of the traditional ``VBM-raising strategy.''
\end{abstract}
%============================================================
%============================================================
\section{Introduction}\label{sec:intro}
%============================================================
%============================================================
Oxide materials are attractive due to
their chemical stability and ease of synthesis under ambient conditions.
In particular, oxide semiconductors are a technologically important class of materials with
wide range of applications in displays~\cite{Nomura2004}, transparent devices~\cite{Pal2024}, power devices~\cite{Higashiwaki_2016},
catalysis~\cite{Baker2020}, and photocatalysis~\cite{KRISHNAN2024389}.
However, a longstanding challenge remains: while several $n$-type oxide semiconductors
(e.g., InGaZnO~\cite{Zhu_2021}, Ga$_2$O$_3$~\cite{Higashiwaki_2016}, ZnO~\cite{SHARMA20223028}, In$_2$O$_3$~\cite{JO2008308},
and SnO$_2$~\cite{Rauf1996}) are well established in consumer electronics,
%==================================
$p$-type oxides are absent from practical applications despite extensive research~\cite{Zhang_2016,Hu2020,Joe2021},
including numerous efforts on doping strategies and device-level demonstrations in applications such as thin-film transistors (TFTs)~\cite{Nomura2004,10.1063/1.2964197,Hosono_2013,LIU201985,Kim2022,Zunger2003,Robertson2021}.
%=======================================
This asymmetry originates from the difficulty of $p$-type doping in oxides,
in which the valence bands are typically dominated by deep O-2\textit{p} orbitals with little dispersion,
leading to higher acceptor formation energies and hole compensation by oxygen vacancies~\cite{Gunkel2020}.

A common strategy for achieving hole dopability in oxides is
to raise the valence band maxima (VBM) by either hybridizing other orbitals with O-2$p$ orbitals
or introducing orbitals energetically above the O-2$p$ levels~\cite{Cao_pTCM,Hu2020,Joe2021,hautier2013identification}.
In this study, we refer to this as the traditional ``VBM-raising strategy.''
For example, CuAlO$_2$, the first $p$-type transparent conducting oxide (TCO),
shows hole conductivity due to strong hybridization between fully occupied Cu-3\textit{d} and O-2\textit{p} states~\cite{kawazoe1997p}.
Similar behavior is observed in other Cu-based delafossites~\cite{pssa200521479,gake2021point,ueda2001epitaxial,coatings9020137}.
The latter case includes SnO~\cite{varley2013ambipolar}, PbO~\cite{C1CS15098G}, BaBiO$_3$~\cite{ShiJueli2022}, TeSeO~\cite{Meng2024} and $\beta$-TeO$_2$~\cite{zavabeti2021high,PhysRevApplied.22.044065,Costa-Amaral2025},
where lone-pair orbitals from cations such as Sn$^{2+}$, Pb$^{2+}$, Bi$^{3+}$ or Te$^{4+}$
lie above the CBM.
The introduction of lone-pair orbitals can also be achieved through alloying, for example, with Bi-doped Ga$_2$O$_3$~\cite{PhysRevB.103.115205}.
The mixture of these two approaches involves introducing partially filled \textit{d} orbitals,
as seen in compounds like NiO~\cite{PhysRevB.85.115127,molaei2013crystallographic}, Cr$_2$O$_3$ \cite{C7TC03545D}, LaVO$_3$~\cite{JELLITE20181}, and V$_2$O$_3$~\cite{Ainabayev2024}.
Many $p$-type TCOs proposed in theoretical studies have also been
suggested based on the VBM-raising strategy~\cite{Zhang_2016,Hu2020,Cao_pTCM}.

While this strategy is well-founded, expanding the effective search for $p$-type oxides remains necessary
because representative $p$-type oxides reported previously exhibit low hole mobilities
(e.g., hole mobilities of 0.25~\mobunit and 18.71~\mobunit in CuAlO$_2$~\cite{LIU2024101304} and SnO~\cite{Caraveo-Frescas2013}, respectively,
vs electron mobilities of 200--250~\mobunit and 125--200~\mobunit in ZnO~\cite{ALFARAMAWI2014} and SnO$_2$~\cite{https://doi.org/10.1002/pssa.201330020}, respectively).
%============================================================
A new avenue for achieving hole doping into O-2p-derived bands is
to focus on the metastable phases.
Indeed, rutile SiO$_2$~\cite{Lyons_2024,chae2025extreme}, cubic ZnO~\cite{PhysRevMaterials.2.084603} and
alloyed BeO with MgO~\cite{chae2025extreme,lyons2025alkali} have been proposed as potential $p$-type oxides.
However, synthesizing these metastable phases remains challenging, which may limit their practical applicability.
%===========================================================

To address this, a precise understanding of the intrinsic difficulty of introducing holes into
O-2$p$-dominated bands is essential.
For example, holes are mostly compensated by charged oxygen vacancies~\cite{PhysRevApplied.19.034063}.
However, our high-throughput calculations on oxygen vacancies reveal that
the formation energies in the +2 charge state span as much as 9.5 eV~\cite{PhysRevMaterials.5.123803}.
This indicates that some oxides may resist hole compensation by the oxygen vacancies due to their high formation energies,
even when the Fermi level lies near the deep VBM composed of the O-2$p$ orbitals.
Therefore, we aim to identify if there are oxides, from 845 candidates, where holes can be introduced into the O-2$p$-dominated bands.
Such insights should provide important design guidelines for the future exploration of $p$-type TCOs.

\begin{figure}
\centering
\includegraphics[width=8cm]{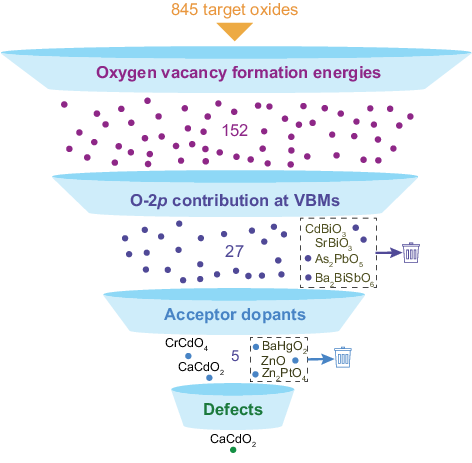}
\captionof{figure}{Flow to screen oxides that can be hole doped into the O-2$p$-dominated bands in this research.
The remaining numbers of oxides in each step are also shown.
The oxides enclosed in the dashed box were excluded from the screening
based on a detailed analysis of their VBM character (see text).
The acceptors calculated in the third step are listed in Table S1 of the Supporting Information, 
and the 152 oxides considered in the second step are listed in Tables S2–S4, categorized according to the O-2$p$ contribution at the VBM.}
\label{fig:flow}
\end{figure}

Our workflow is illustrated in Fig.~\ref{fig:flow}.
First, we screen oxides based on the formation energies of oxygen vacancies.
Next, we quantify the orbital contributions to the VBM of the screened oxides
and retain those whose VBM are primarily composed of the O-2$p$ orbitals.
We then discuss the acceptor levels introduced by cationic dopants in the remaining oxides.
Finally, we evaluate the formation energies of native defects and acceptor dopants
in the screened oxides using the HSE06 hybrid functional,
in order to investigate the hole dopability of the O-2$p$-dominated bands.
We here note that the the sequence of the screening steps does not affect the results,
provided that the same criteria are applied.
%-----------------------------------------
%---------------------------------------
%============================================================
%============================================================
\section*{Results and discussion}
%============================================================
%============================================================
\subsection*{Oxygen vacancy formation energies}
%============================================================
%============================================================
%-----------------------------------------
\begin{figure}
\centering
\includegraphics[width=16cm]{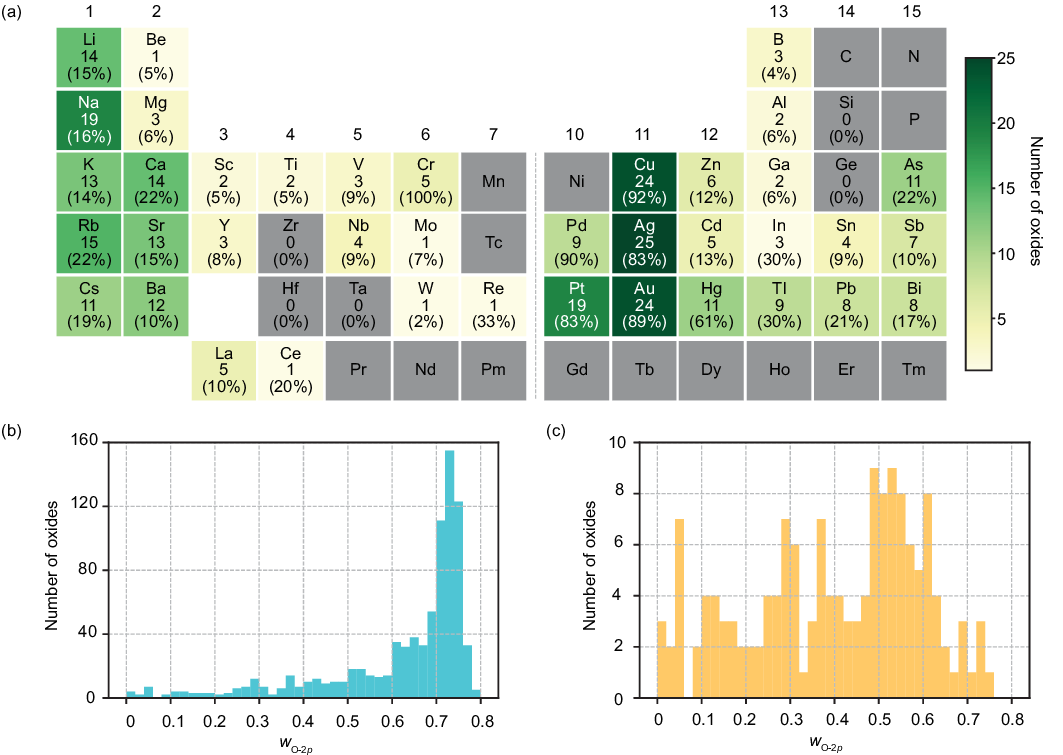}
\captionof{figure}{
(a) Number of oxides containing each element among the 152 oxides retained after screening based on oxygen vacancy formation energies.
 The percentages in parentheses indicate the survival rates after screening.
 The number of oxides containing each element before the screening are shown in \si.
(b, c) Histogram of the O-2$p$ contribution at the VBM ($w_{\mathrm{O\text{-}2}p}$) for (b) 845 target oxides
 and (c) 152 oxides remaining after screening.
 }
\label{fig:statistial}
\end{figure}
%---------------------------------------
Since oxygen vacancies are major ``hole killers,''
screening oxides by their formation energies efficiently filters out unsuitable oxides.
The targets are the same 845 oxides as those in Ref.~\cite{PhysRevApplied.19.034063},
which satisfy the following criteria:
(i) stable against competing phases,
(ii) band gaps are larger than 0.3 eV in the Materials Project Database (MPD),
(iii) nonmagnetic,
and
(iv) the number of atoms in their primitive cells is 30 or less in the MPD
(See Method for details).
Note that, as mentioned above, while metastable phases can be promising $p$-type oxides,
we exclude them from this study to focus on candidates that are experimentally easily accessible under ambient conditions.
Based on these criteria, SnO is excluded from our targets because it is unstable with respect to SnO$_2$ and O$_2$.
To avoid complex systems comprising mixed anions and/or cations with spin-polarized $d$ or $f$ electrons,
oxides containing H, He, C–Ne, P–Ar, Mn–Ni, Se–Kr, Tc–Rh, Te–Xe, Pr–Lu, Os, Ir, and Po–Lr were excluded.

When a sufficient number of holes are introduced, the Fermi level is positioned near the VBM.
For example, in ZnO, when the hole concentration is 10$^{17}$ cm$^{-3}$,
the Fermi level lies at 0.2 eV at 300~K (see \si).
Therefore, to prevent hole compensation by oxygen vacancies,
all inequivalent oxygen vacancies must be stable near the VBM, either
(i) in neutral charge states or
(ii) in positive charge states with positive formation energies.
In this study, we considered the O-rich conditions to suppress oxygen vacancies as much as possible (see Method).

Consequently, only 152 out of the 845 oxides satisfy this criterion.
They are primarily composed of group 10 and 11 elements (Pd, Pt, Cu, Ag, Au) (Fig.~\ref{fig:statistial}(a)),
which tend to possess fully occupied $d$ orbitals.
In contrast, oxides containing early transition-metal cations (e.g., Sc, Y, Ti, V, Nb) are relatively scarce,
mainly because their unoccupied $d$ states do not typically contribute to raising the O-$2p$ manifold.
Oxides containing alkali and alkaline earth metals remain relatively abundant,
as they are components of ternary or quaternary oxides.

%============================================================
\subsection*{O-2$p$ orbital contributions at valence band maxima}
%============================================================
\begin{figure*}
\centering
\includegraphics[width=17cm]{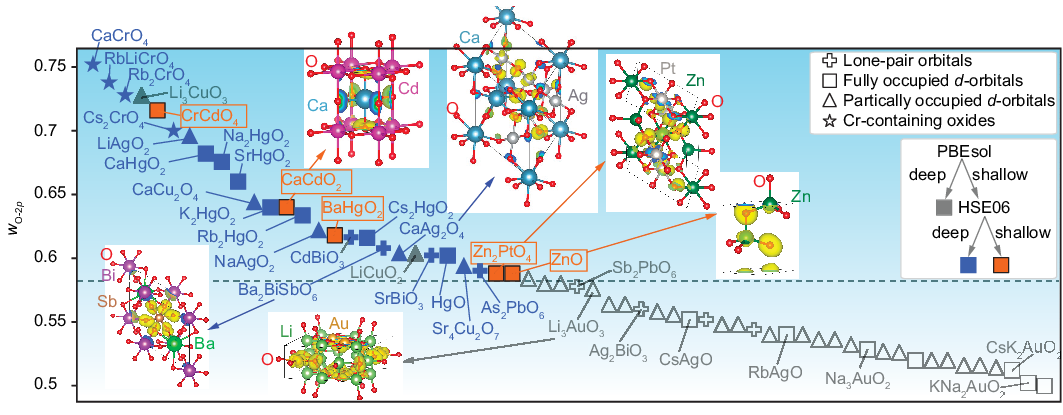}\\
\captionof{figure}{
 The O-2$p$ contributions at the VBM ($w_{\mathrm{O\text{-}2}p}$) are shown in descending order.
 Oxides are categorized into four classes, as indicated in the legend (see text for details).
 For oxides with $w_{\mathrm{O\text{-}2}p}$ greater than that of ZnO, the acceptor levels introduced by dopants
 are also classified into three categories.
 Gray symbols indicate oxides with deep acceptor levels in PBEsol calculations.
 Blue symbols represent oxides that have shallow acceptor levels in PBEsol but become deep in HSE06.
 Orange symbols denote oxides with shallow acceptor levels even when using HSE06.
 Isosurface plots of representative oxides are shown at 5\% of the maximum electron density.
}
\label{fig:o2pcontribution}
\end{figure*}
% ---------------------------------------

To identify oxides whose VBM mainly consists of O–2$p$ orbitals,
we computed the projected values of the VBM wavefunctions onto
the O-2$p$ orbitals ($w_{\mathrm{O\text{-}2}p}$) for 845 oxides (see Method).
As shown in Fig.~\ref{fig:statistial} (b),
most oxides exhibit substantial $w_{\mathrm{O\text{-}2}p}$, typically above 0.6.
After screening based on the oxygen vacancy formation energies,
the average $w_{\mathrm{O\text{-}2}p}$ value decreases from 0.63 (Fig.~\ref{fig:statistial}(b)) to 0.40 (Fig.~\ref{fig:statistial}(c)).
This highlights that oxides whose VBMs are mainly composed of O–2$p$ orbitals are largely excluded.

Figure~\ref{fig:o2pcontribution} shows $w_{\mathrm{O\text{-}2}p}$ and the corresponding names of oxides,
sorted in descending order of $w_{\mathrm{O\text{-}2}p}$.
Here, we classify oxides into four categories.
The first category includes oxides with lone-pair orbitals near the VBM,
namely Pb$^{2+}$ and Bi$^{3+}$.
The second and third categories consist of oxides with fully or partially occupied $d$ electrons
in or near the valence bands, respectively.
In the former case, the oxides include Ag$^{+}$, Au$^{+}$, Zn$^{2+}$, Cd$^{2+}$, and Hg$^{2+}$,
whereas in the latter case, they include Cu$^{3+}$, Ag$^{3+}$, Pt$^{4+}$, and Pd$^{4+}$.
The remaining compounds include Cr-containing oxides, namely CaCrO$_4$, RbLiCrO$_4$, Rb$_2$CrO$_4$, and Cs$_2$CrO$_4$.
When an oxide falls into more than one category, it is assigned to the earliest applicable category.
Such oxides are CrCdO$_4$, Ag$_2$BiO$_3$, CdBiO$_3$, CdAuO$_2$ (Au$^{+}$ and Au$^{3+}$), and Rb$_5$Au$_3$O$_2$ (Au$^{-}$),
the last two of which are classified as oxides with fully occupied $d$ electrons.
All the Cr-containing oxides exhibit significantly high $w_{\mathrm{O\text{-}2}p}$,
indicating a lower contribution of unoccupied Cr-3$d$ orbitals to the VBM.
A detailed electronic structures of Cr-containing oxides are discussed later.

%============================

In Figure~\ref{fig:o2pcontribution},
we show isosurfaces of squared wavefunctions at the VBM in representative oxides (see Method).
These illustrate how $w_{\mathrm{O\text{-}2}p}$ relates to the spatial distributions of the VBM.
Oxides in the middle-to-lower region of Figure~\ref{fig:o2pcontribution}
tend to exhibit hybridization of O-2$p$ orbitals with other orbitals.
For example, the O-2$p$ orbitals in ZnO are slightly hybridized with the deep Zn-3$d$ orbitals.
Ba$_2$BiSbO$_6$ and CaAg$_2$O$_4$ also show contributions from Bi lone pairs and Ag-4$d$ orbitals, respectively.
On the other hand, CaCdO$_2$, which exhibits a higher $w_{\mathrm{O\text{-}2}p}$ of 0.64, shows only O-2$p$ character.
However, it should be noted that a higher $w_{\mathrm{O\text{-}2}p}$ does not always correspond to
a dominant O-2$p$ character at the VBM, as exemplified by Ba$_2$BiSbO$_6$,
where the presence of Bi lone-pair orbitals is evident.
This indicates that analyzing the VBM character using $w_{\mathrm{O\text{-}2}p}$ is qualitative.
Thus, we applied a looser criterion by retaining oxides with $w_{\mathrm{O\text{-}2}p}$
greater than that of ZnO (0.588) to avoid false negatives;
As a result, 27 out of 152 oxides remained.
We also manually excluded CdBiO$_3$, SrBiO$_3$, As$_2$PbO$_5$, and Ba$_2$BiSbO$_6$
because they exhibit lone-pair orbital characteristics as exemplified by Ba$_2$BiSbO$_6$ in Figure~\ref{fig:o2pcontribution} and \si.
Consequently, 23 oxides were retained for further analysis.
Note that CuAlO$_2$ ($w_{\mathrm{O\text{-}2}p}$=0.128) and the well-known $p$-type B$_6$O~\cite{Akashi2002} ($w_{\mathrm{O\text{-}2}p}$=0.118)
are excluded at this step, since their VBMs are mainly composed of Cu-3$d$ and B-2$p$ orbitals~\cite{PhysRevB.90.045205}, respectively.
%%===================
Some representative $p$-type TCOs that satisfy the ``VBM-raising strategy''
are listed in the Table S5 in \si,
together with the steps where they are excluded from our screening.
This supports the validity of our screening.
%
%============================================================
\subsection{Acceptor doping}\label{subsec:acceptor-dopants}
%============================================================
When synthesizing $p$-type semiconductors, the presence of dopants that exhibit shallow acceptor levels is crucial.
Therefore, in the next step, we calculated the acceptor dopants in the remaining 23 oxides.
In this study, we considered acceptor-type dopants substituting at cationic sites,
as this type of doping can be readily achieved by simply sintering the host oxide
together with a dopant-containing oxide during synthesis.
%We selected representative acceptors: Na for divalent, Ca for trivalent,
%Sc for tetravalent, Si for pentavalent, and V for hexavalent cations.
%=======================
We selected representative acceptor dopants whose ionic radii are closest to those of the host cations: 
Li, Na, or K for divalent; Ca or Zn for trivalent; Sc or Al for tetravalent; and V for hexavalent cations. 
The details of the calculated dopants and their corresponding ionic radii are provided in Table S1.
%===============================
Note that, with this approach, oxides composed solely of monovalent cations are not hole-dopable,
since it is generally uncommon to dope closed-shell noble gas elements.
Fortunately, none of the remaining 23 oxides fall into this category.

As shown in Figure~\ref{fig:o2pcontribution}, most oxides (21 out of 23)
are predicted to exhibit shallow acceptor levels based on PBEsol calculations.
This number is, however, significantly reduced when using the more accurate HSE06 hybrid functional,
which provides improved band gap estimations,
leaving only five oxides: CrCdO$_4$, CaCdO$_2$,  BaHgO$_2$, Zn$_2$PtO$_4$, and ZnO.
This result indicates that employing more accurate functionals than PBEsol is crucial for reliably identifying hole-dopability in oxides.
Unfortunately, although Cr-containing oxides exhibit higher $w_{\mathrm{O\text{-}2}p}$,
all of them exhibit deep acceptor levels except for CrCdO$_4$ (see \si).

Electronic band structures, densities of states (DOS), and squared wavefunctions at the VBM
for the remaining five oxides are shown in Figure~\ref{fig:banddos}(a--e).
They confirm that the O-2$p$ orbitals predominantly compose the valence bands in all these compounds.
However, all of them exhibit fully occupied $d$ semi-core states between $-8$ and $-5$~eV relative to each VBM,
which are expected to slightly hybridize with the O-2$p$ orbitals, tending to elevate the VBM.
Indeed, the squared wavefunctions of BaHgO$_2$, ZnO, and Zn$_2$PtO$_4$ reveal the presence Zn-3$d$ or Hg-5$d$ orbital components,
despite their relatively high $w_{\mathrm{O\text{-}2}p}$ values.
Thus, we conclude that hole doping into the pure O-2$p$-dominated bands is nearly impossible,
and at least such a small hybridization with the valence band is
necessary to enable hole doping into the O-2$p$ orbitals.

The electronic band structures and DOS for CaCrO$_4$, RbLiCrO$_4$, Rb$_2$CrO$_4$, and Cs$_2$CrO$_4$
are shown in Figure S9 in \si.
Similar to CrCdO$_4$, all of them exhibit valence bands dominated by O-2$p$ states.
Interestingly, RbLiCrO$_4$, Rb$_2$CrO$_4$, and Cs$_2$CrO$_4$ show very low valence band dispersion.
This is primarily due to the low oxygen ion density in these oxides.
Consequently, only CrCdO$_4$, which is classified as a fully occupied $d$-electron oxide,
shows higher hybridization of the VBM with the Cr-3$d$ orbitals, and thus a shallow acceptor level.

We emphasize that ZnO remains in our screening.
Indeed, Tsukazaki \etal reported \textit{p}-type ZnO using nitrogen as a $p$-type dopant~\cite{Tsukazaki2005},
demonstrating a light-emitting diode; however, hole doping in ZnO remains extremely challenging~\cite{PhysRevB.81.205209,10.1063/1.3274043,PhysRevB.80.085202}.
See the recent review paper~\cite{Yang2023}, which outlines various stringent conditions for realizing \textit{p}-type ZnO.

%============================================================
\subsection{Defect formation energies in CrCdO$_4$ and CaCdO$_2$}\label{sec:defect-formation-energies-in-crcdo$_4$-and-cacdo$_2$}
%============================================================
%----------------------------------------
\begin{figure*}
\centering
\includegraphics[width=17cm]{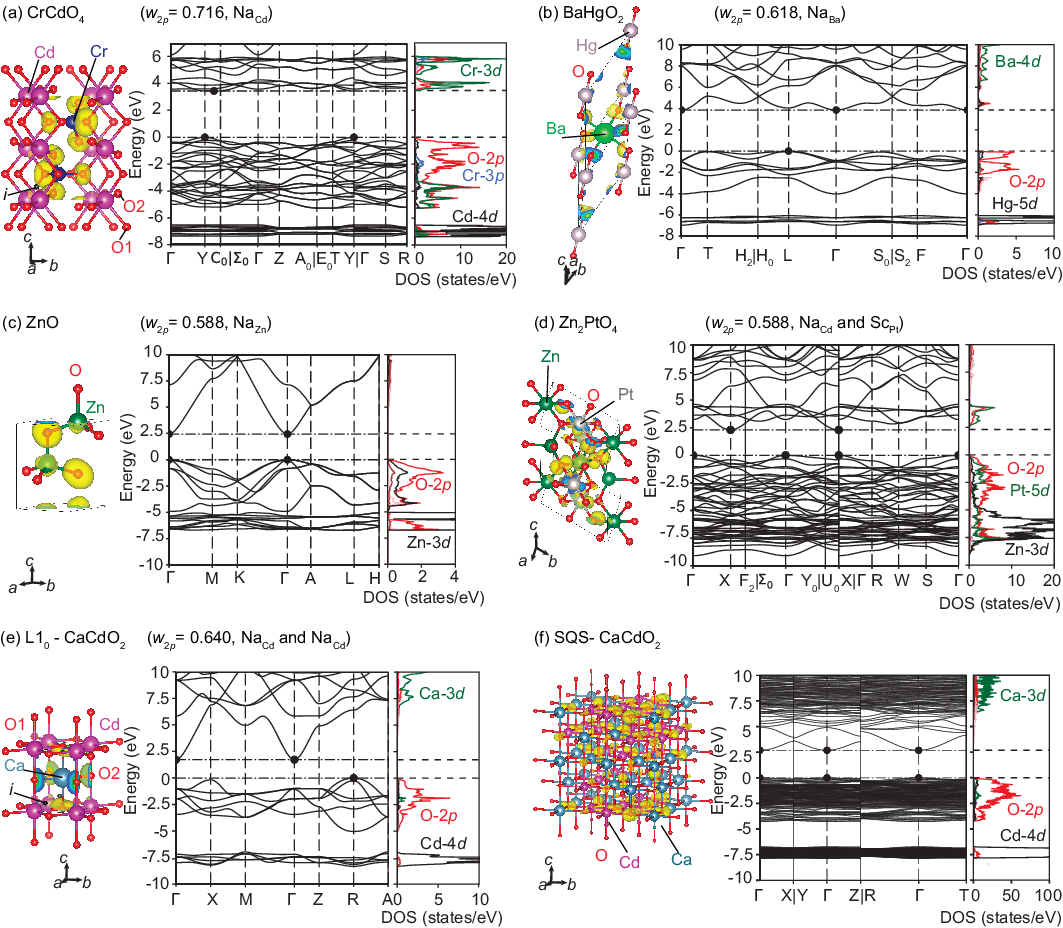}\\
\captionof{figure}{
    Crystal structures with squared wavefunctions at the VBM, electronic band structures, and densities of states for
    (a) CrCdO$_4$, (b) BaHgO$_2$, (c) ZnO, (d) Zn$_2$PtO$_4$, (e) L1$_0$--CaCdO$_2$, and (f) SQS model for CaCdO$_2$.
    The O-2$p$ contributions at the VBM ($w_{\mathrm{O\text{-}2}p}$) and the dopants confirmed to behave as shallow acceptors are also indicated in parentheses.
}
\label{fig:banddos}
\end{figure*}
% ---------------------------------------
\begin{figure*}
\centering
\includegraphics[width=17cm]{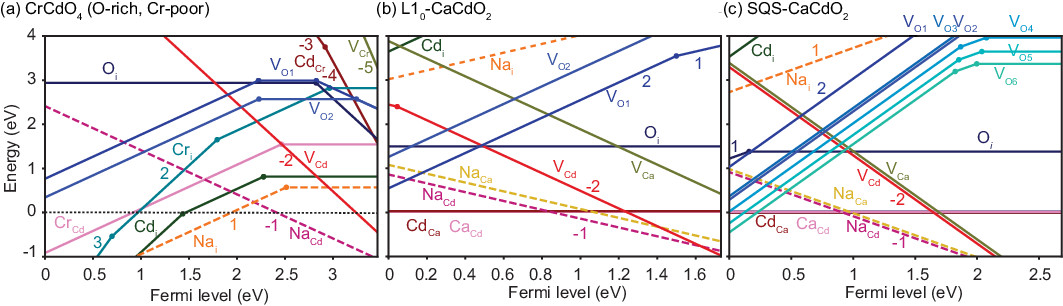}\\
\captionof{figure}{
    Formation energies of native defects (solid lines) and Na dopants (dashed lines)
    as a function of the Fermi level in
    (a) CrCdO$_4$ (O-rich, Cr-poor), (b) L1$_0$--CaCdO$_2$, and (c) SQS model for CaCdO$_2$.
    The chemical potentials are set at oxygen-rich conditions (see \si).
    $V_X$ denotes a vacancy at the $X$ site, while $X_i$ denotes an interstitial at the $X$ position.
    The atomic sites and interstitial sites are shown in the crystal structures in Figure~\ref{fig:banddos}.
    $X_Y$ denotes an antisite defect where $X$ is substituted at the $Y$ site.
    In (c), only the lowest formation energies are shown for each defect type except for the oxygen vacancies.
    }
\label{fig:formation}
\end{figure*}
% ---------------------------------------

Since CrCdO$_4$ and CaCdO$_2$ exhibit stronger O-2$p$ character in Figure~\ref{fig:banddos},
it is of importance whether they can exhibit hole conductivity.
To determine this, we need the total energies of all point defects
that may compensate carrier holes and/or pin the Fermi level, as well as the acceptor dopants.
We therefore performed their calculations using the HSE06 hybrid functional.
We consider vacancies, cation antisites, and interstitials as native defects, and
acceptor-type dopants substituted at cation or interstitial sites (see Methods).

%CrCdO$_4$
CrCdO$_4$ crystallizes in the orthorhombic space group $Cmcm$.
Experimental reports are very limited:
its crystal structure was refined by Müller \etal~\cite{Muller1969},
and Bodade \etal measured the frequency-dependent dielectric constant of Sr-doped CrCdO$_4$~\cite{Bodade2015}.
Based on our HSE06 calculations, CrCdO$_4$ exhibits relatively low hole effective masses
of 1.35, 1.73, and 1.64~$m_0$ along the $x$-, $y$-, and $z$-directions, respectively (see \si),
where $m_0$ is the electron rest mass.
These values indicate promising hole mobility.
The calculated optical band gap is 3.48~eV,
which suggests that it could serve as a $p$-type TCO if sufficient hole doping is achievable.

The formation energies of native defects under O-rich and Cd-poor conditions are shown in Fig.~\ref{fig:formation}(a).
Oxygen vacancies have positive formation energies even when the Fermi level is near the VBM, consistent with our initial screening.
However, interstitials generally have lower formation energies, which significantly restricts the accessible Fermi level range.
As a result, achieving a high hole concentration would be difficult.

%CaCdO$_2$
The structure of CaCdO$_2$ registered in the MPD belongs to the space group $P4/mmm$,
in which cations are arranged in the L1$_0$-type layered configuration.
However, CaCdO$_2$ is known to exhibit cation disorder experimentally,
forming a cation disordered rock-salt structure with space group $Fm\bar{3}m$.
This behavior can be rationalized by the identical oxidation states
and similar ionic radii of Ca$^{2+}$ (1.00~\AA) and Cd$^{2+}$ (0.95~\AA) in sixfold coordination~\cite{shannon1976revised}.
Experimental data for CaCdO$_2$ are also scarce, with the notable exception of the study by Srihari et al.~\cite{10.1063/1.3526300},
which reported how the lattice constant and band gap vary as a function of Ca fraction.

We modeled the disordered phase using the special quasi-random structure (SQS) technique~\cite{VANDEWALLE201313} (see Method).
The atomic arrangement, band structure, DOS,
and squared wavefunction at the VBM of the SQS model are shown in Figure~\ref{fig:banddos}(f).
The fundamental gap of the ordered model is 1.73~eV, while that for the SQS model is 2.68~eV.
This large discrepancy is primarily due to the narrower dispersion of the conduction band in the SQS model;
in both the ordered and SQS models, the conduction bands are composed of the Ca-3$d$ orbitals,
and their interactions are weaker in the SQS model due to the inhomogeneous bonding between Ca atoms.
The optical gap of the SQS model is 3.30~eV, which is higher than the experimental value of 2.83~eV~\cite{10.1063/1.3526300}.
Since such an optical gap is nearly transparent, CaCdO$_2$ is also expected to be a \emph{p}-type TCO.
%The hole effective masses in the SQS model are about 1.8--1.9~$m_0$,
%which is comparable to those in GaN (1.40$\pm$0.33~$m_0$)~\cite{PhysRevB.62.7365}.
%===============================
The conventional band structure is not strictly
defined for a SQS, due to the lack of translational symmetry. However, in this
study, we employed relatively small SQS supercells.
In practice, effective masses estimated from such small SQS supercells remain
useful, as the associated band dispersion reflects the density of states, which
in turn can be used to evaluate the effective masses via effective density of
states. 
The SQS model yields nearly isotropic hole effective masses in all
directions ($\sim$1.8–1.9$m_0$), which are slightly larger than that of GaN (1.40$\pm$0.33$m_0$)~\cite{PhysRevB.62.7365}  and similar to that in in-plane of CuAlO$_2$ ($\sim$1.9$m_0$)~\cite{PhysRevMaterials.3.044603}.
%=====================================

Focusing on the point defects in the ordered CaCdO$_2$ (Fig.~\ref{fig:formation}(b)),
the oxygen vacancies again exhibit positive formation energies under the O-rich and $p$-type conditions, consistent with our screening.
However, unlike CrCdO$_4$, other donor-type defects such as cation interstitials exhibit higher formation energies,
and do not introduce any pinning levels.
This would be simply because the vacant space in the rock-salt structure is smaller than CrCdO$_4$.
In addition, Na dopants exhibit lower formation energies at the Cd sites (Na$_\mathrm{Cd}$)
and higher formation energies at the interstitial sites,
which hinders the self-compensation of Na dopants.
As anticipated from the disordering behavior, cation antisites exhibit very low formation energies,
but they are stable in the neutral charge states and thus do not affect the carrier concentration.
In CaO, oxygen interstitials exhibit amphoteric behavior and low formation energies,
introducing deep pinning levels that prohibit a high concentration of hole doping~\cite{10.1063/5.0211707}.
However, this is not the case in CaCdO$_2$, due to its higher VBM compared to CaO.
Consequently, a high hole concentration can be introduced by Na dopants (see~\si).

%sqs-CaCdO2
Since experimental samples of CaCdO$_2$ are disordered, 
defect formation energies computed using the SQS model are particularly relevant. 
We investigated cation vacancies, cation antisites, and Na substitution at three randomly selected Ca and three Cd sites (see \si), 
based on the fact that each cation site is coordinated by six nearest-neighbor oxygen atoms and the site dependence of formation energies is relatively small. 
In contrast, oxygen vacancies were analyzed according to the number of surrounding Cd atoms, 
since oxygen vacancy formation energies in CdO and CaO are nearly identical despite the VBM of CdO being 1.5~eV higher (see Fig.~\ref{fig:alignment} shown later),
suggesting weaker Cd--O bonding compared to Ca--O. 
Indeed, as Cd coordination increases, the formation energy of oxygen vacancies tends to decrease (see Fig.~\ref{fig:formation}(c)). 
Interestingly, except for the oxygen vacancies, the formation energies of point defects in the ordered CaCdO$_2$ and SQS model are almost the same
when we assume the VBM positions are close to each other (see Fig.~\ref{fig:formation}(b,c)).

%===========================
Although CaCdO$_2$ emerges as the most promising candidate,
it is desirable to replace Cd with less toxic elements.
A common strategy is to replace toxic elements with non-toxic ones from the same group,
    such as substituting Pb with Sn.
In group 12, possible substitutions for Cd include Zn and Hg;
    however, Hg is also toxic, and replacing Cd with Zn is challenging since Zn favors a coordination number of four.
For instance, stabilizing rocksalt-type ZnO at ambient pressure is nearly impossible~\cite{PhysRevMaterials.2.084603}.
%
%============================================================
\subsection{Band alignment}\label{subsec:discussion}
%============================================================
\begin{figure}
\centering
\includegraphics[width=8cm]{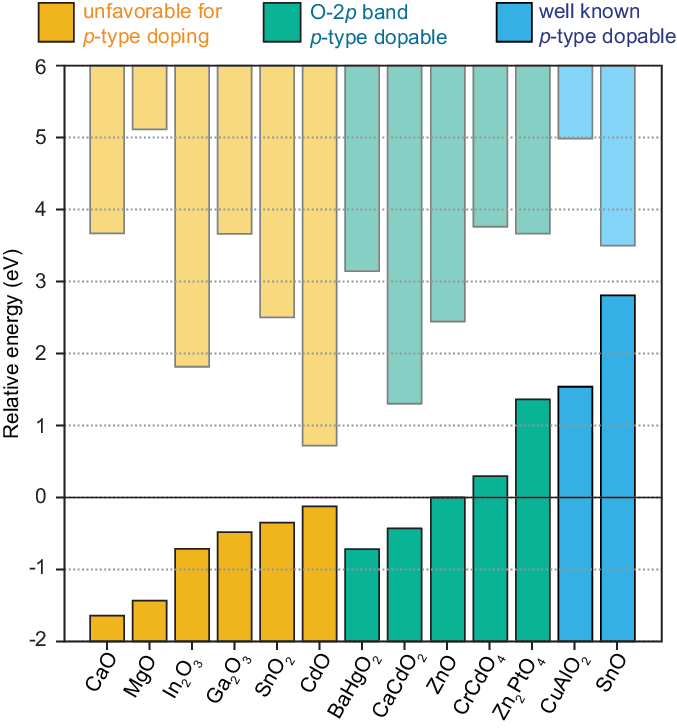}\\
\captionof{figure}{
    Band alignment obtained using HSE06, showing the VBM across three groups:
(i) oxides unfavorable for $p$-type doping, 
(ii) oxides identified in this study as O-2$p$-dominated band $p$-type dopable candidates, 
and (iii) well-known $p$-type dopable oxides.
The energy zero is set to the VBM in ZnO.
}
\label{fig:alignment}
\end{figure}

Finally, we discuss hole dopability based on band alignment.
Several references can be used for aligning band edges,
such as the vacuum level estimated from slab models~\cite{PhysRevB.95.125309},
branch point energies (BPE)~\cite{10.1063/1.3059569},
and core potentials~\cite{PhysRevLett.44.1620,PhysRevB.28.1965,10.1116/1.584941}.
We note that none of these methods is perfect.
For example, using the vacuum level, the alignment depends on the surface orientation,
while the BPE largely depends on the number of bands considered~\cite{10.1063/1.3059569,Woods-Robinson2018} (see~\si).
In this study, we constructed the band alignment using the O-2$p$ core potentials.
The advantage of this approach is that it eliminates such arbitrariness;
however, it is a bold assumption that the oxygen core potential remains constant across all oxides.
For reference, we also present the band alignment based on the BPE in \si.

Figure~\ref{fig:alignment} presents the band alignment obtained from HSE06 calculations for:
(i) oxides unfavorable for $p$-type doping;
(ii) oxides identified in this study as hole-dopable candidates
into the O-2$p$ orbitals by cationic acceptor dopants; and
(iii) CuAlO$_2$ and SnO, which satisfy the ``VBM-raising strategy.''
CuAlO$_2$ and SnO, well-known $p$-type TCOs, exhibit significantly higher VBMs,
and Zn$_2$PtO$_4$ and CrCdO$_4$ also show relatively higher VBM compared to hole undopable oxides.
%=======================
From Figure~6, we observe that oxides with a large O--2$p$ contribution tend to exhibit low VBMs, 
such as CaO ($w_{\mathrm{O\text{-}2}p}$= 0.705), MgO ($w_{\mathrm{O\text{-}2}p}$ = 0.747), Ga$_2$O$_3$ ($w_{\mathrm{O\text{-}2}p}$= 0.750), and SnO$_2$ ($w_{\mathrm{O\text{-}2}p}$ = 0.768). 
In contrast, the VBM of CuAlO$_2$ is significantly higher due to the strong contribution from Cu-3$d$ orbitals, while its O-2$p$ contribution is relatively small ($w_{\mathrm{O\text{-}2}p}$ = 0.128).
%====================================
However, the hole dopability into O-2$p$ orbitals appears difficult to predict only from the alignment,
as illustrated by the comparison between CaCdO$_2$ and CdO.
One of the primary reasons is that the band alignment does not include information on defect formation energies,
which depend on the controllability of the chemical potential.
Previous studies have used band alignment to screen materials for dopability~\cite{PRXEnergy.1.033006,Yim2018-op,Woods-Robinson2018}.
Based on our results, caution is needed to avoid screening out many false negatives.

%============================================================
\section{Conclusions}\label{sec:conclusions}
%============================================================
In this work, we systematically screened stable 845 oxides to identify those
in which holes can be introduced into the VBM mainly composed of O-2$p$ orbitals.
Our results underscore the intrinsic challenge of hole doping in these systems,
with only CaCdO$_2$ emerging from the screening.
Notably, its VBM is slightly hybridized with deep-lying Cd-3$d$ orbitals.
This highlights that, without cation $d$ orbitals or lone-pair bands,
hole doping into a pure O-2$p$-dominated band is extremely difficult, if not impossible.
Thus, our research reinforces the effectiveness of the ``VBM-raising strategy.''
We also investigated whether hole dopability can be simply predicted from the band alignment,
and found that while a high VBM is related to hole dopability, it is insufficient alone,
as dopability is also influenced by other factors such as the controllability of the chemical potential.

Finally, we emphasize the limitations of our screening approach.
There are two main limitations:
First, we focus only on thermodynamically stable oxides at 0~K in the Materials Project.
As described earlier, some metastable oxides have been proposed as candidates for ambipolar or $p$-type doping.
We set this limitation because we aim to identify $p$-type oxides that can be readily synthesized in experiments.
Second, we assessed dopability solely based on whether the acceptor dopants exhibit hydrogenic states.
Then,  potential candidates may be overlooked if dopants induce deep localized states
    whose transition levels are nevertheless relatively shallow.
However, we emphasize that, in practice, such defects are unlikely to enable effective hole doping,
    as exemplified by the long-standing difficulty of achieving \textit{p}-type GaN~\cite{RevModPhys.87.1133,10.1063/5.0022198}.
Thus, our study does not claim that CaCdO$_2$ is the only candidate for a $p$-type oxide with O-2$p$-dominated valence bands.
We also note that the effective masses associated with the O-2$p$-dominated band
    are generally large due to its relatively localized nature
    although there are some exceptions (see Fig. S15 in Supporting Information).
    Therefore, even if hole doping into the O-2$p$-dominated band is achieved,
    the oxide may still be unsuitable for practical applications.

%============================================================
\section{Methods}\label{sec:method}
%============================================================
%============================================================
\subsection{Computational setting of first-principles calculations}\label{subsec:computational}
%============================================================
%First-principles calculations
First-principles calculations were performed
using the projector augmented-wave (PAW) method~\cite{blochl1994projector,kresse1999ultrasoft},
as implemented in VASP~\cite{PhysRevB.54.11169}.
Details of the PAW potentials are provided in Ref.~\cite{PhysRevApplied.19.034063,PhysRevMaterials.5.123803}.
In this study, we used the PBEsol exchange-correlation functional~\cite{Schimka2011}
and the standard HSE06 hybrid functional~\cite{krukau2006influence}.
Effective mass tensors were computed using BoltzTraP2~\cite{MADSEN200667,madsen2018boltztrap2},
assuming a typical carrier concentration of $10^{18}~\mathrm{cm}^{-3}$ and a temperature of 300~K.
For the optical absorption spectra,
the real parts of the dielectric functions
were obtained from the imaginary parts via the Kramers–Kronig transformation,
using a complex shift of 0.01~eV~\cite{madsen2018boltztrap2}.
The optical band gaps were then defined from the Tauc plots (see \si).
Band paths for electronic structure calculations were generated automatically
using SeeK-path~\cite{hinuma2017band}.
The values of the VBM wavefunctions projected onto the O-2$p$ orbitals ($w_{\mathrm{O\text{-}2}p}$)
were calculated using the oxygen PAW core radius of 0.804~\AA.
We constructed the band alignment using oxygen core potentials as a reference.
When there are multiple O sites, we used the averaged value.
Crystal structures and isosurface plots were generated using VESTA~\cite{Momma:db5098}.
We constructed the 64-atom SQS model using the ATAT code~\cite{VANDEWALLE2009266}.
Starting from a $2\times 2\times 2$ conventional rock-salt structure,
we included clusters up to quadruplets and interactions up to the second-nearest neighbors.
All VASP input files were generated using VISE~\cite{PhysRevMaterials.5.123803},
and defect models were constructed and analyzed with pydefect~\cite{PhysRevMaterials.5.123803}.
More details, such as parameter settings, are described in \si.

%============================================================
\subsection{Target oxides}
Since we utilize the previously calculated oxygen vacancy formation energy database,
we used the same 845 target oxides as in Ref.~\cite{PhysRevApplied.19.034063}.
The target oxides were retrieved on May 14, 2021, from the Materials Project Database (MPD)~\cite{jain2013commentary}.
Note that their stability was determined based on the PBE(+U) calculations at that time~\cite{Momma:db5098}.

%============================================================
\subsection{Modeling of point defects}\label{subsec:defect}
%============================================================
% Defect calculations 
%
Defect calculations were carried out using supercells containing 192 atoms for CrCdO$_4$,
and 64 atoms for both ordered and disordered CaCdO$_2$ (see \si).
The supercells used for calculating the acceptor dopants in the other oxides are described in \si.
The internal atomic positions were relaxed while keeping the lattice constants fixed at their theoretical values.
In the initial defect structures, small displacements to the neighboring atoms were introduced
to break the symmetry and explore lower-energy configurations.
All defect calculations allowed spin polarization.
The interstitial sites were determined from the local minima of the electron charge density, as shown in Fig.~\ref{fig:banddos} (a) and (f).
The oxygen interstitial sites in CaCdO$_2$ are found to be stable in split-type O$_2$ dimers,
as plotted in Figure S11(b) and S12(d) in \si.
In Section~\nameref{subsec:acceptor-dopants}, 
we validated the shallow acceptor behavior from single-particle levels (see \si).
The carrier concentrations as a function of the Fermi level were calculated
from the DOS via the Fermi-Dirac distribution~\cite{PhysRevB.90.125202}.
The defect concentrations were estimated from the Boltzmann distribution,
considering the site degeneracies that depend on the defect site symmetries in the relaxed structures
and the spin degeneracies~\cite{PhysRevB.90.125202}, and shown in Figure S3 in \si.
The synthesis temperatures for each candidate were set
based on the conditions reported in previous experiments~\cite{Tsukazaki2005,Yang2023,THAOWANKAEW201814172,MIKHAILOVA1999151,Bodade2015,10.1063/1.3526300}.

% Dielectric constant and finite-size corrections
The finite-size effects on the formation energies of charged defects were corrected using
the extended FNV method~\cite{PhysRevLett.102.016402,PhysRevB.89.195205,PhysRevB.90.125202}.
Static dielectric tensors ($\epsilon_{0}$) were obtained
as the sum of the high-frequency ion-clamped dielectric tensor ($\epsilon_{\infty}$)
and the ionic contribution ($\epsilon_{\mathrm{ion}}$),
calculated using density functional perturbation theory~\cite{PhysRevB.33.7017,PhysRevB.73.045112,PhysRevB.63.155107}.
$\epsilon_{\infty}$ were derived from HSE06 calculations,
while $\epsilon_{\mathrm{ion}}$ were computed using PBEsol to reduce computational cost.
Local field effects were included in the evaluation of $\epsilon_{\infty}$.
Dielectric constants of the SQS model were approximated by taking the average of those in CdO and CaO,
since their calculations are computationally too demanding.
% Chemical potential diagrams
Chemical potential diagrams (CPDs) were constructed using the total energies calculated using HSE06 at 0~K.
The energy difference between the ordered and disordered structures of CaCdO$_2$ is only 0.3~meV/atom.
Since they are slightly unstable with respect to competing phases,
we decreased their total energies by only 5~meV/atom so that they appear in the CPD.
Candidate competing phases were obtained from the Materials Project~\cite{jain2013commentary}.
A full list of competing phases used is provided in \si.

%============================================================
%============================================================
\begin{suppinfo}
The Supporting Information is available free of charge at  [TO BE INSERTED].
Additional computational results and data, including the inverse average hole effective masses as a function of the band gap for 845 oxides, elemental distribution among target oxides, carrier concentrations versus Fermi level, isosurface plots of lone-pair oxides, single-particle dopant levels (HSE06), electronic band structures and densities of states, chemical-potential diagrams and full defect formation energies for CrCdO$_4$ and CaCdO$_2$, band-edge alignment and Tauc plots, and comparison of O-2$p$ contributions at the VBM obtained with different PAW potentials. 
Tables summarize dopant selections and ionic radii, screened oxide candidates with Materials Project-IDs and categories, representative $p$-type TCOs from literature, anisotropic hole effective masses, chemical potentials, competing phases, defect-site coordinates, and equilibrium Fermi levels with carrier and defect concentrations. 
\end{suppinfo}
%
%============================================================
\begin{acknowledgement}
This work has been supported by JST FOREST Program (JPMJFR235S), and KAKENHI (22H01755 and 25K01486); 
the E-IMR project at IMR, Tohoku University; and the Center for Diversity, Equity, and Inclusion (TUMUG) at
Tohoku University, through the Project to Promote Gender Equality and Female Researchers. 
Some calculations were conducted using MASAMUNE-IMR supercomputers.
\end{acknowledgement}
%============================================================
%============================================================
%============================================================
%============================================================
%
%\bibliographystyle{naturemag}
%\bibliography{oxides.bib}
\providecommand{\latin}[1]{#1}
\makeatletter
\providecommand{\doi}
  {\begingroup\let\do\@makeother\dospecials
  \catcode`\{=1 \catcode`\}=2 \doi@aux}
\providecommand{\doi@aux}[1]{\endgroup\texttt{#1}}
\makeatother
\providecommand*\mcitethebibliography{\thebibliography}
\csname @ifundefined\endcsname{endmcitethebibliography}
  {\let\endmcitethebibliography\endthebibliography}{}

\appendix


\section*{Supplementary material}

Table S1: List of 27 oxides with $w_{\text{O-}2p}$ higher than that of ZnO (0.588), screened as candidates for $p$-type oxides in this study, based on oxygen vacancy formation energies. Their Materials Project ID (MP-ID), space group, $w_{\text{O-}2p}$, category as defined in this study (see main text), acceptor dopants considered, and the number of atoms in the supercell for defect calculations ($N_{\text{atoms}}$) are also shown. The categories that contain "Partially (Full) occupied $d$-orbitals" are abbreviated as "Partially (Full)" for simplicity. **Bolded** doping defects indicate those that survived as shallow acceptors under HSE06. Since $CdBiO_3$, $SrBiO_3$, $As_2PbO_5$, and $Ba_2BiSbO_6$ exhibit lone-pair orbital characteristics, their dopant calculations were not performed. Oxides containing only monovalent cations and lone-pair electrons are excluded (see text).

| MP-ID | Formula | Space group | $w_{\text{O-}2p}$ | Categories | Doping | $N_{\text{atoms}}$ |
|---|---|---|---|---|---|---|
| mp-19215 | $CaCrO_4$ | $I4_1/amd$ | 0.752 | Cr-based | $Na_{Ca}$, $V_{Cr}$ | 192 |
| mp-18741 | $RbLiCrO_4$ | $P3_1c$ | 0.738 | Cr-based | $V_{Cr}$ | 112 |
| mp-17071 | $Rb_2CrO_4$ | $Pnma$ | 0.728 | Cr-based | $V_{Cr}$ | 56 |
| mp-19970 | $Li_3CuO_3$ | $P4_2/mnm$ | 0.728 | Partially | $Ca_{Cu}$ | 234 |
| mp-18781 | $CrCdO_4$ | $Cmcm$ | 0.716 | Fully | **$Na_{Cd}$**, $V_{Cr}$ | 192 |
| mp-647813 | $Cs_2CrO_4$ | $Pnma$ | 0.700 | Cr-based | $V_{Cr}$ | 224 |
| mp-996962 | $LiAgO_2$ | $P2_1/c$ | 0.696 | Partially | $Ca_{Ag}$ | 288 |
| mp-7041 | $CaHgO_2$ | $R\bar{3}m$ | 0.682 | Fully | $Na_{Hg}$, $Na_{Ca}$ | 300 |
| mp-4961 | $Na_2HgO_2$ | $I4/mmm$ | 0.676 | Fully | $Na_{Hg}$ | 160 |
| mp-665922 | $SrHgO_2$ | $R\bar{3}m$ | 0.660 | Fully | $Na_{Hg}$, $Na_{Sr}$ | 300 |
| mp-1047437 | $CaCu_2O_4$ | $I4_1/a$ | 0.644 | Partially | $Na_{Ca}$, **$Ca_{Cu}$** | 224 |
| mp-5860 | $K_2HgO_2$ | $I4/mmm$ | 0.640 | Fully | $Na_{Hg}$ | 160 |
| mp-753287 | $CaCdO_2$ | $P4/mmm$ | 0.640 | Fully | **$Na_{Cd}$**, **$Na_{Ca}$** | 64 |
| mp-5072 | $Rb_2HgO_2$ | $I4/mmm$ | 0.632 | Fully | $Na_{Hg}$ | 160 |
| mp-754326 | $NaAgO_2$ | $C2/m$ | 0.622 | Partially | $Ca_{Ag}$ | 128 |
| mp-3915 | $BaHgO_2$ | $R\bar{3}m$ | 0.618 | Fully | **$Na_{Ba}$**, $Na_{Hg}$ | 300 |
| mp-755478 | $CdBiO_3$ | $R3$ | 0.616 | Lone-pair | – | – |
| mp-7232 | $Cs_2HgO_2$ | $I4/mmm$ | 0.616 | Fully | $Na_{Hg}$ | 160 |
| mp-23127 | $Ba_2BiSbO_6$ | $C2/m$ | 0.608 | Lone-pair | – | – |
| mp-9158 | $LiCuO_2$ | $C2/m$ | 0.604 | Partially | $Ca_{Cu}$ | 128 |
| mp-1047036 | $CaAg_2O_4$ | $I4_1/a$ | 0.604 | Partially | $Ca_{Ag}$, $Na_{Ca}$ | 122 |
| mp-29164 | $SrBiO_3$ | $P2_1/c$ | 0.602 | Lone-pair | – | – |
| mp-1224 | $HgO$ | $Pnma$ | 0.602 | Fully | $Na_{Hg}$ | 96 |
| mp-766217 | $Sr_4Cu_2O_7$ | $P2_1/c$ | 0.594 | Partially | $Na_{Sr}$, $Ca_{Cu}$ | 104 |
| mp-20015 | $As_2PbO_6$ | $P\bar{3}1m$ | 0.590 | Lone-pair | – | – |
| mp-35647 | $Zn_2PtO_4$ | $Imma$ | 0.588 | Fully | **$Na_{Zn}$**, **$Sc_{Pt}$** | 122 |
| mp-2133 | $ZnO$ | $P6_3mc$ | 0.588 | Fully | **$Na_{Zn}$** | 300 |

2

Table S2: Similar to Table S1 but for 33 oxides having $w_{\text{O-}2p}$ smaller than that of ZnO and greater than 0.5. These oxides were only checked for shallow acceptor levels using PBEsol.

| MP-ID | Formula | Space group | $w_{\text{O-}2p}$ | Categories | Doping | $N_{\text{atoms}}$ |
|---|---|---|---|---|---|---|
| mp-4541 | NaCuO$_2$ | $C2/m$ | 0.584 | Partially | Ca$_{\text{Cu}}$ | 64 |
| mp-997088 | KAgO$_2$ | $Cmcm$ | 0.580 | Partially | Ca$_{\text{Ag}}$ | 128 |
| mp-1096933 | CsAgO$_2$ | $Cmcm$ | 0.580 | Partially | Ca$_{\text{Ag}}$ | 128 |
| mp-20727 | Sb$_2$PbO$_6$ | $P\bar{3}1m$ | 0.578 | Lone-pair | – | – |
| mp-7471 | Li$_3$AuO$_3$ | $P4_2/mnm$ | 0.572 | Partially | Ca$_{\text{Au}}$ | 224 |
| mp-768915 | Na$_3$AuO$_3$ | $P4_2/mnm$ | 0.564 | Partially | Ca$_{\text{Au}}$ | 224 |
| mp-782050 | Ba$_4$PdO$_6$ | $R\bar{3}c$ | 0.564 | Partially | Na$_{\text{Ba}}$, Sc$_{\text{Pd}}$ | 66 |
| mp-23558 | Ag$_2$BiO$_3$ | $Pnna$ | 0.560 | Lone-pair | – | – |
| mp-1041631 | Mg(CuO$_2$)$_2$ | $Pbcm$ | 0.558 | Partially | Na$_{\text{Mg}}$, Ca$_{\text{Cu}}$ | 224 |
| mp-997052 | RbAgO$_2$ | $Cmcm$ | 0.556 | Partially | Ca$_{\text{Ag}}$ | 128 |
| mp-8666 | CsAgO | $I4/mmm$ | 0.552 | Fully | – | – |
| mp-22984 | NaBiO$_2$ | $C2/c$ | 0.552 | Lone-pair | – | – |
| mp-1095378 | AsAuO$_4$ | $C2/c$ | 0.548 | Partially | Si$_{\text{As}}$, Sc$_{\text{Au}}$ | 194 |
| mp-553310 | CsCuO$_2$ | $Cmcm$ | 0.548 | Partially | Ca$_{\text{Cu}}$ | 128 |
| mp-22169 | MoPbO$_4$ | $P\bar{1}$ | 0.544 | Lone-pair | – | – |
| mp-29775 | Sr$_4$PdO$_6$ | $R\bar{3}c$ | 0.540 | Partially | Na$_{\text{Sr}}$, Sc$_{\text{Pd}}$ | 66 |
| mp-8603 | RbAgO | $I4/mmm$ | 0.540 | Fully | – | – |
| mp-3982 | KCuO$_2$ | $Cmcm$ | 0.540 | Partially | Ca$_{\text{Cu}}$ | 96 |
| mp-10299 | Ca$_4$PdO$_6$ | $R\bar{3}c$ | 0.536 | Partially | Na$_{\text{Ca}}$, Sc$_{\text{Pd}}$ | 66 |
| mp-28796 | Sr$_5$(AuO$_4$)$_2$ | $I4/mmm$ | 0.536 | Partially | Ca$_{\text{Au}}$ | 120 |
| mp-997002 | KAuO$_2$ | $Cmcm$ | 0.532 | Partially | Ca$_{\text{Au}}$ | 96 |
| mp-28365 | Na$_3$AuO$_2$ | $P4_2/mnm$ | 0.528 | Fully | – | – |
| mp-27253 | Au$_2$O$_3$ | $Fdd2$ | 0.527 | Partially | Ca$_{\text{Au}}$ | 120 |
| mp-997089 | NaAuO$_2$ | $Cmcm$ | 0.526 | Partially | Ca$_{\text{Au}}$ | 96 |
| mp-29920 | Rb$_5$Au$_3$O$_2$ | $Pbam$ | 0.520 | Fully | Ca$_{\text{Au}}$ | 120 |
| mp-7472 | RbAuO$_2$ | $Cmcm$ | 0.520 | Partially | Ca$_{\text{Au}}$ | 96 |
| mp-997024 | CsAuO$_2$ | $Cmcm$ | 0.520 | Partially | Ca$_{\text{Au}}$ | 96 |
| mp-8727 | Ba$_4$PtO$_6$ | $R\bar{3}c$ | 0.516 | Partially | Na$_{\text{Ba}}$, Sc$_{\text{Pt}}$ | 66 |
| mp-757242 | Li$_5$AuO$_4$ | $Fddd$ | 0.516 | Partially | Ca$_{\text{Au}}$ | 160 |
| mp-760483 | Li$_2$PdO$_3$ | $C2/m$ | 0.514 | Partially | Sc$_{\text{Pd}}$ | 96 |
| mp-559067 | KNa$_2$AuO$_2$ | $Pnnm$ | 0.512 | Fully | – | – |
| mp-557807 | CsK$_2$AuO$_2$ | $Pnnm$ | 0.502 | Fully | – | – |
| mp-997036 | CdAuO$_2$ | $P\bar{1}$ | 0.500 | Fully | Na$_{\text{Cd}}$ | 96 |

Table S3: List of 92 oxides having $w_{\text{O-}2p}$ smaller than 0.5. Acceptor dopants for these oxides were not calculated.

| Formula | $w_{\text{O-}2p}$ | Formula | $w_{\text{O-}2p}$ | Formula | $w_{\text{O-}2p}$ |
| --- | --- | --- | --- | --- | --- |
| $RbNa_2AuO_2$ | 0.498 | $Tl_3BO_3$ | 0.366 | $La_2Pd_2O_5$ | 0.196 |
| $CsAuO$ | 0.496 | $Ba_4AgAuO_6$ | 0.364 | $LiCuO$ | 0.184 |
| $Sr_4PtO_6$ | 0.496 | $LiAgO$ | 0.356 | $CsCu_3O_2$ | 0.180 |
| $Sr(AuO_2)_2$ | 0.488 | $CaAgAsO_4$ | 0.350 | $Cu_2PbO_2$ | 0.178 |
| $Ba(AuO_2)_2$ | 0.488 | $Ca_2Pt_3O_8$ | 0.344 | $Ba_2Ca(PdO_2)_3$ | 0.172 |
| $RbAuO$ | 0.488 | $Na_2ReO_3$ | 0.340 | $Sr(CuO)_2$ | 0.160 |
| $Na_3CuO_2$ | 0.488 | $TiSnO_3$ | 0.320 | $CsCuO$ | 0.160 |
| $LiAuO_2$ | 0.482 | $Tl_3AsO_4$ | 0.316 | $CuAsPbO_4$ | 0.156 |
| $KNa_2CuO_2$ | 0.480 | $LiAg_3O_2$ | 0.316 | $NaCuO$ | 0.132 |
| $Mg_2PtO_4$ | 0.472 | $NaAg_3O_2$ | 0.312 | $AlCuO_2$ | 0.128 |
| $Na_2PtO_3$ | 0.464 | $Ag_2HgO_2$ | 0.304 | $Ba(CuO)_2$ | 0.124 |
| $Sr_3CdPtO_6$ | 0.460 | $Li_2PdO_2$ | 0.302 | $GaCuO_2$ | 0.124 |
| $Sr_3ZnPtO_6$ | 0.452 | $Na_3AgO_2$ | 0.300 | $KCuO$ | 0.120 |
| $Tl_2Pt_2O_7$ | 0.452 | $AgAuO_2$ | 0.300 | $B_6O$ | 0.118 |
| $Li_8PtO_6$ | 0.446 | $NaAgO$ | 0.296 | $TlPd_3O_4$ | 0.112 |
| $Rb_3CuO_2$ | 0.440 | $AlAgO_2$ | 0.288 | $TiNb_3O_6$ | 0.108 |
| $Rb_3SbO_3$ | 0.433 | $BeTlAsO_4$ | 0.288 | $Zn(PtO_2)_3$ | 0.098 |
| $Sr_3MgPtO_6$ | 0.420 | $Rb_2Sn_2O_3$ | 0.285 | $LaCuO_2$ | 0.096 |
| $RbCuO$ | 0.420 | $LiNbO_2$ | 0.284 | $TlPt_3O_4$ | 0.056 |
| $Ca(AuO_2)_2$ | 0.410 | $Hg_2WO_4$ | 0.276 | $LaZnAsO$ | 0.050 |
| $In_2Pt_2O_7$ | 0.408 | $GaAgO_2$ | 0.272 | $Ca_4As_2O$ | 0.046 |
| $Cs_2Pb_2O_3$ | 0.402 | $TlSbO_3$ | 0.264 | $La_3SbO_3$ | 0.044 |
| $K_2PdO_2$ | 0.398 | $K_2Sn_2O_3$ | 0.264 | $YZnAsO$ | 0.044 |
| $Ba_2CePtO_6$ | 0.396 | $CaNb_2O_4$ | 0.256 | $Ca_4Sb_2O$ | 0.042 |
| $Rb_2Pb_2O_3$ | 0.390 | $NaNbO_2$ | 0.256 | $Ca_4Bi_2O$ | 0.041 |
| $La_2Pt_2O_7$ | 0.388 | $KAgO$ | 0.244 | $Sr_4As_2O$ | 0.035 |
| $Na_2PtO_2$ | 0.380 | $Tl_2SnO_3$ | 0.242 | $Sr_4Bi_2O$ | 0.023 |
| $K_2Pb_2O_3$ | 0.378 | $ScAgO_2$ | 0.228 | $RbBa_4Sb_3O$ | 0.000 |
| $Y_2Pt_2O_7$ | 0.376 | $Tl_4V_2O_7$ | 0.225 | $Ba_3BAsO_3$ | 0.000 |
| $Sc_2Pt_2O_7$ | 0.372 | $VAg_3HgO_4$ | 0.217 | $KBa_4Bi_3O$ | 0.000 |
| $VAg_3O_4$ | 0.372 | $YAgO_2$ | 0.206 | | |

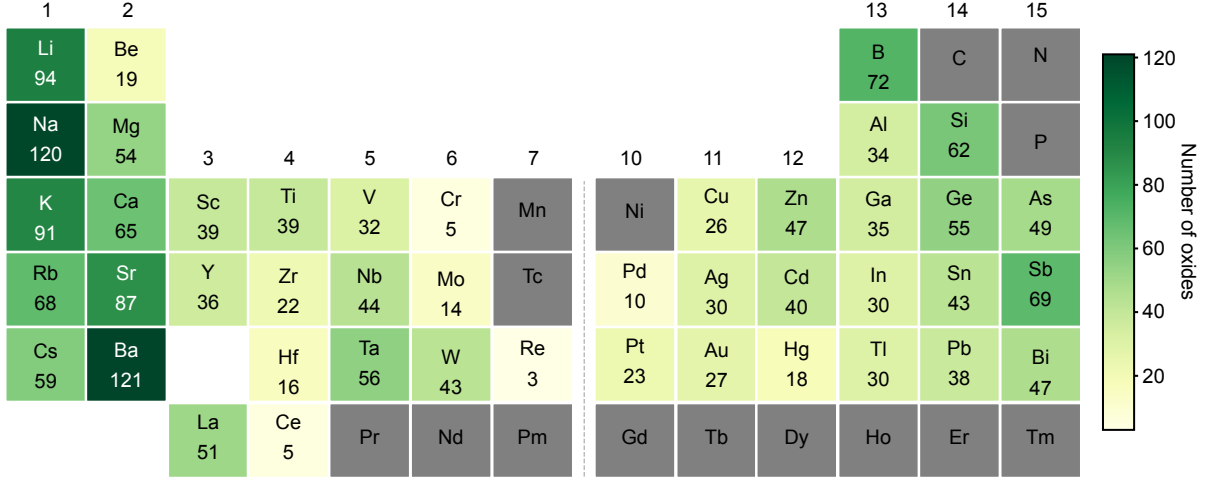

Figure S1: Number of oxides containing each element among the target 845 oxides.

Table S4: Representative *p*-type TCOs from the literature and the step at which each is excluded in this study are summarized. The Te containing oxides are excluded in the first step to avoid mixed-anion oxides. We then check the stability, oxygen vacancy formatio energies, and the O-2*p* contribution at the VBM. Details are described in the main text.

|  | Step to be excluded in this work | | | | | |
| --- | --- | --- | --- | --- | --- | --- |
| Oxides | Containing Te | Stability | $E_{f[V_O]}$ | $w_{\text{O-}2p} < 0.588$ | $w_{\text{O-}2p}$ | References |
| $La_2O_2Te$ | ✓ | | | | — | 1 |
| $\beta$–$TeO_2$ | ✓ | | | | — | 2 |
| SnO | | ✓ | | | — | 3 |
| $Sb_4Cl_2O_5$ | | ✓ | | | — | 4 |
| CuLiO | | | ✓ | | — | 1 |
| $CuAlO_2$ | | | | ✓ | 0.128 | 5 |
| $B_6O$ | | | | ✓ | 0.118 | 6–8 |
| $K_2Sn_2O_3$ | | | | ✓ | 0.264 | 4 |
| $K_2Pb_2O_3$ | | | | ✓ | 0.378 | 4,8 |
| $Na_3AgO_2$ | | | | ✓ | 0.300 | 8 |
| $Rb_2Pb_2O_3$ | | | | ✓ | 0.390 | 8 |
| $Cs_2Pb_2O_3$ | | | | ✓ | 0.402 | 8 |
| CsCuO | | | | ✓ | 0.160 | 1,8 |

5

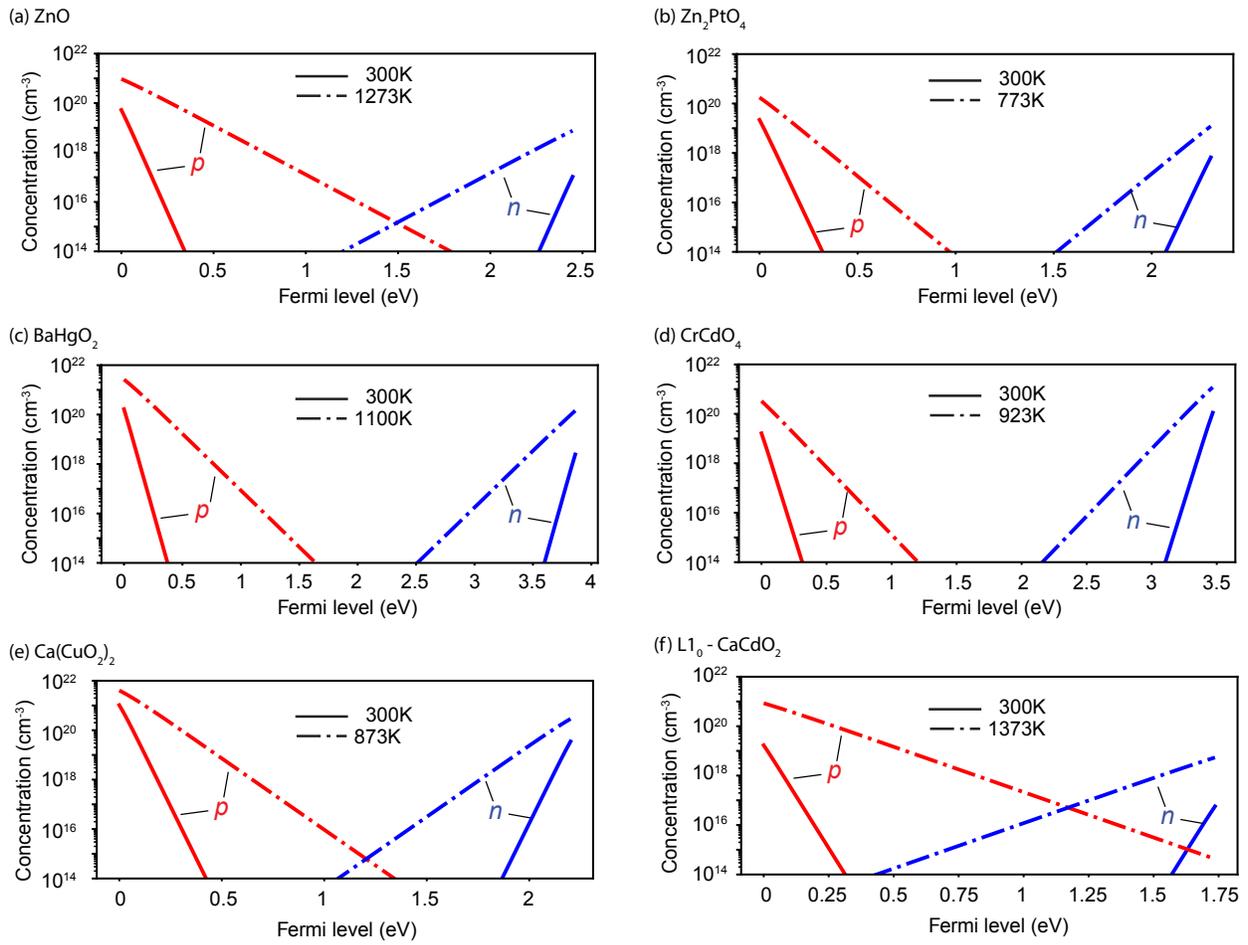

Figure S2: Carrier concentrations as a function of the Fermi level.

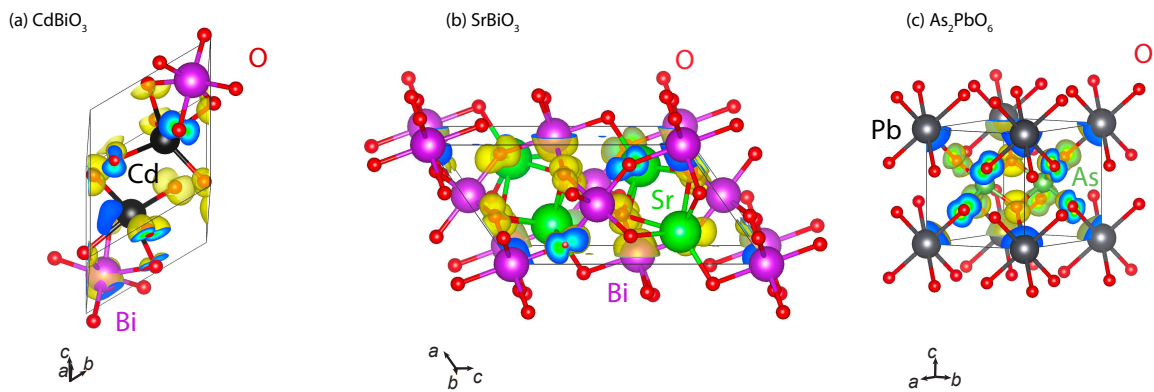

Figure S3: Isosurface plots of representative lone-pair oxides, each shown at 5% of the maximum electron density.

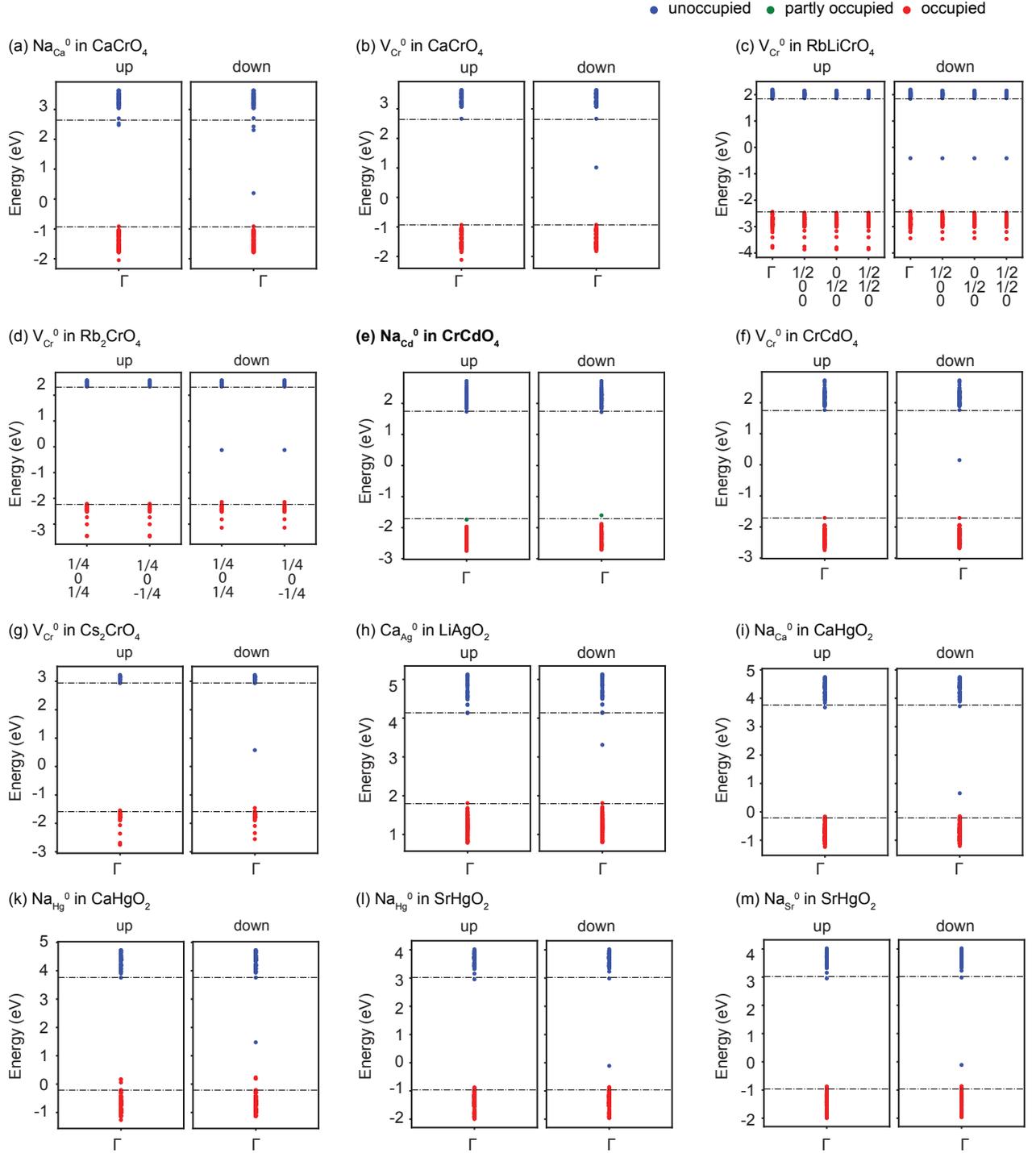

Figure S4: Single-particle levels of shallow acceptor dopants calculated using HSE06 for oxides that show shallow acceptor levels by dopants with PBEsol. **Bold** doping defects indicate those that remain shallow acceptors using HSE06. Both spin-up and spin-down levels are displayed for spin-polarized systems, while unpolarized systems are shown in a single panel. The vertial axis shown with absolute values.

7

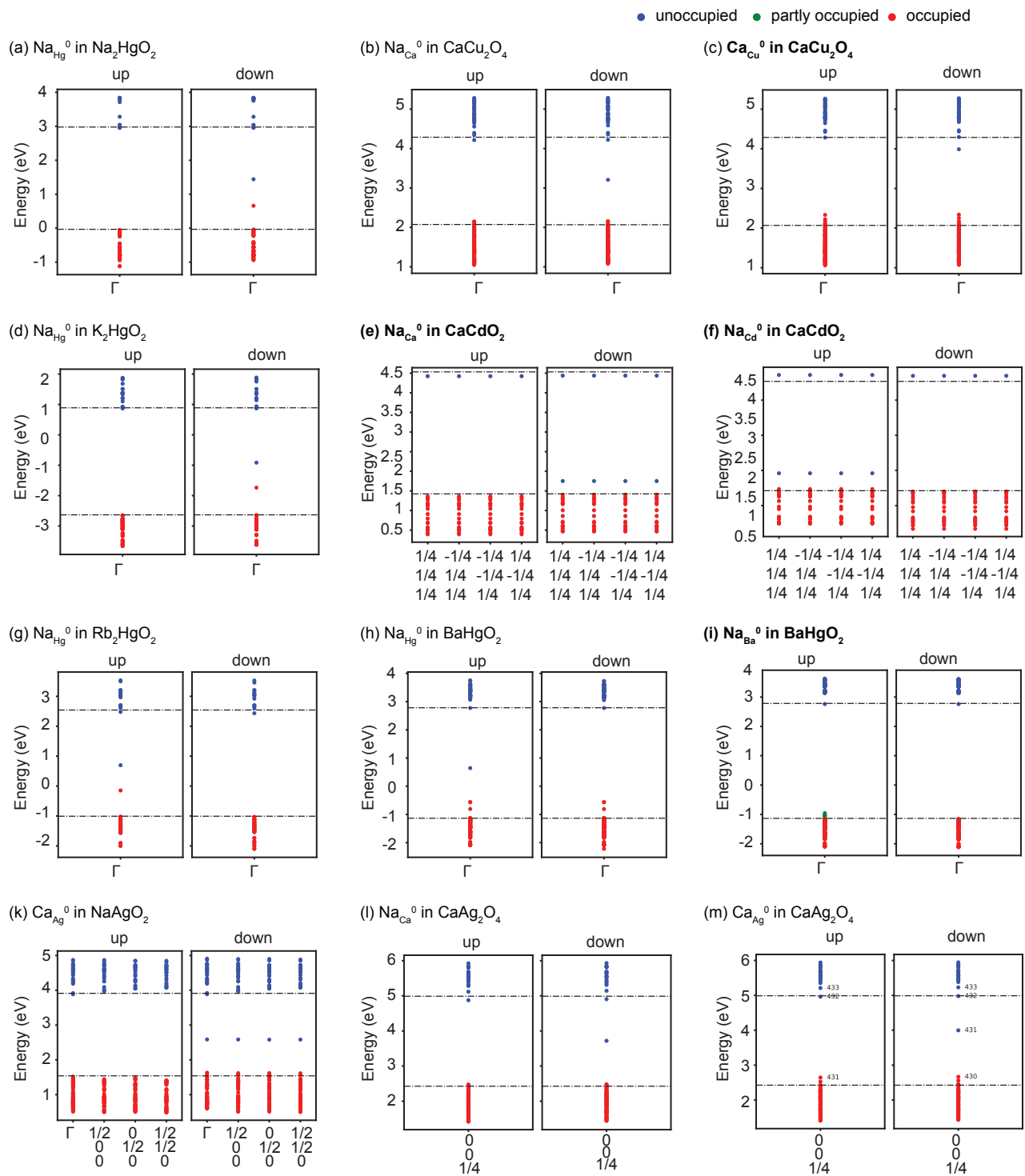

Figure S5: Continuation of Fig. S4.

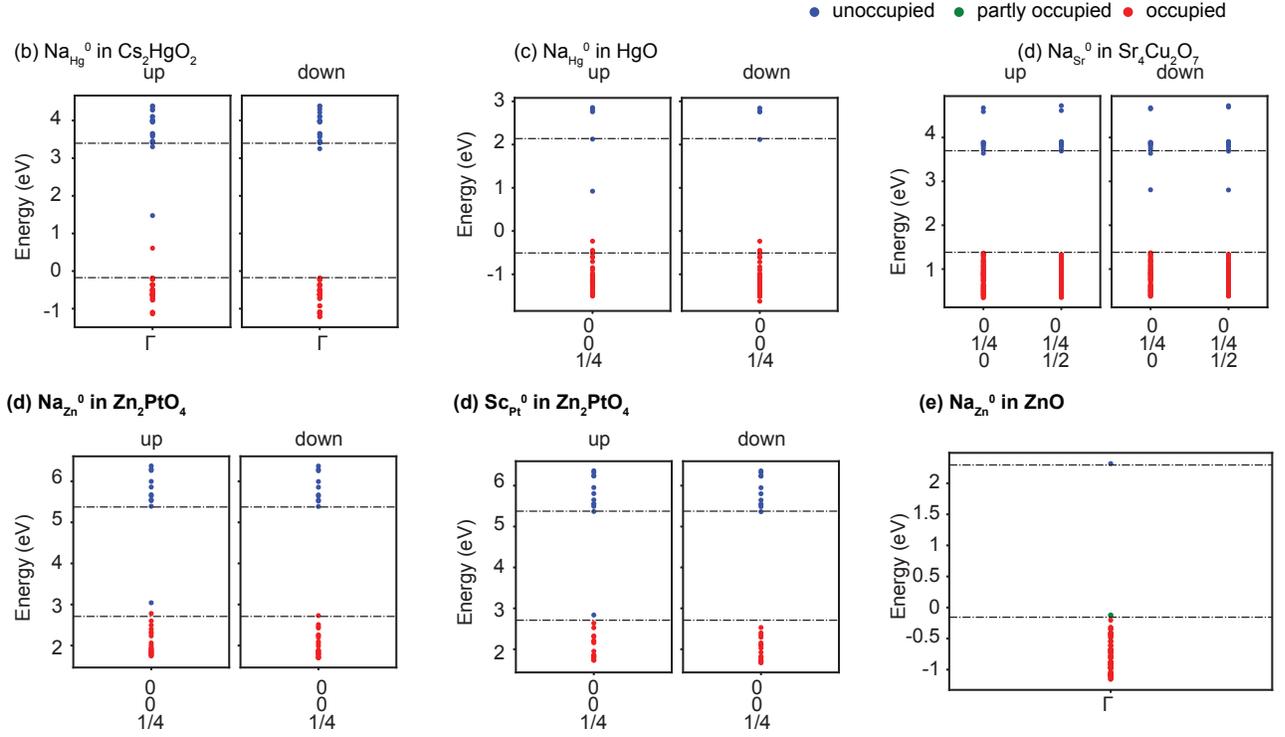

Figure S6: Continuation of Fig. S5.

Table S5: Crystal systems and hole effective masses of six candidates and SQS model.

| Compound | Crystal system | $m^*_{h,x}\ (m_0)$ | $m^*_{h,y}\ (m_0)$ | $m^*_{h,z}\ (m_0)$ |
|---|---|---|---|---|
| ZnO | hexagonal | 1.76 | 1.76 | 2.57 |
| CrCdO$_4$ | orthorhombic | 1.35 | 1.73 | 1.64 |
| BaHgO$_2$ | rhombohedral | 2.47 | 2.47 | 1.06 |
| Zn$_2$PtO$_4$ | orthorhombic | 0.83 | 4.22 | 1.83 |
| CaCu$_2$O$_4$ | tetragonal | 8.54 | 8.55 | 3.33 |
| CaCdO$_2$ | tetragonal | 1.62 | 1.62 | 0.42 |
| sqs–CaCdO$_2$ | cubic | 1.83 | 1.88 | 1.83 |

9

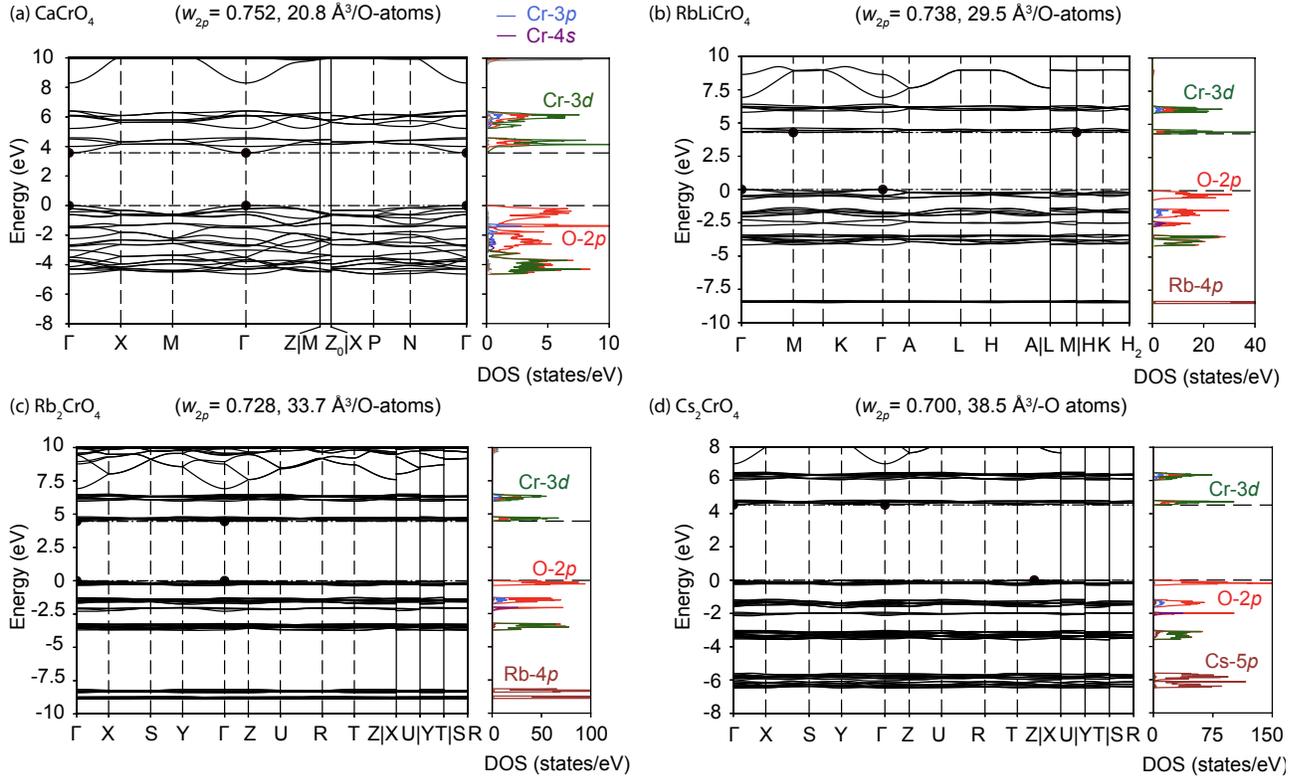

Figure S7: Electronic band structures and densities of state for four Cr-containing oxides. The O-2$p$ contributions at the VBM ($w_{\text{O-2}p}$) and the unitcell volume per O-atoms are also shown in parentheses.

Table S6: Chemical potentials ($\Delta\mu_i$, where $i$ means the impurity elements) of the ternary systems that include each impurity at points A (O-rich, Cr-poor), B, C (O-rich, Cd-poor), and D in CPD in Fig. S8 (a) for CrCdO$_4$. The values are relative to those of the standard states, namely simple substances or molecules. The competing phases that include impurity elements are described.

|  | A-point | B-point | C-point | D-point |
| --- | --- | --- | --- | --- |
| $\Delta\mu_{\text{Cd}}$ | -2.11 | -0.68 | -4.17 | -3.84 |
| $\Delta\mu_{\text{Cr}}$ | -6.72 | -2.43 | -4.66 | -4.33 |
| $\Delta\mu_{\text{O}}$ | 0.0 | -1.43 | 0.0 | -0.16 |
| $\Delta\mu_{\text{Na}}$ | -3.17 | -2.46 | -5.71 | -5.39 |
| Competing phase | CdO, O$_2$(g) | CdO, Cr$_2$O$_3$ | CrO$_2$, O$_2$(g) | Cr$_2$O$_3$, CrO$_2$ |
| Impurity phase | Na$_2$CrO$_4$ | Na$_2$CrO$_4$ | NaCr$_3$O$_8$ | NaCr$_3$O$_8$ |

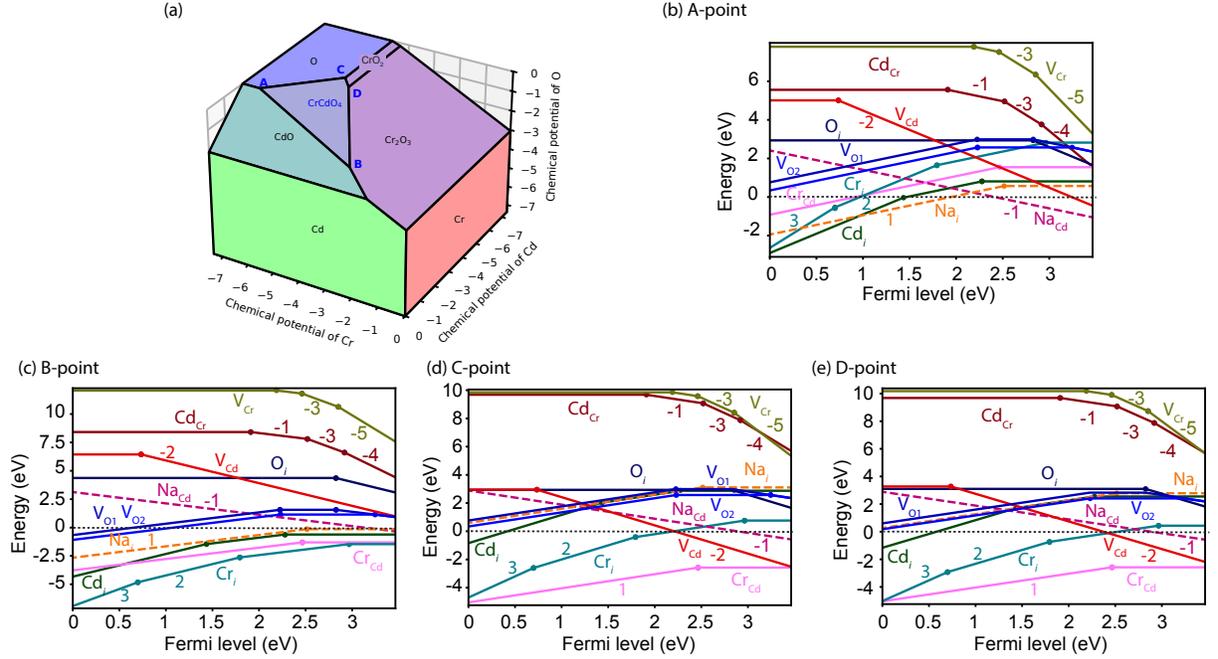

Figure S8: (a) Chemical potential diagram of Cr-Cd-O ternary system. (b-e) Formation energies of native defects and Na dopants in $CrCdO_4$ as function of the Fermi level at A, B, C, D chemical potentials.

Table S7: Same as Table. S6 but for points A, B, C, and D in CPD in Fig. S9 (a) for $CaCdO_2$.

|  | A-point | B-point | C-point | D-point |
| --- | --- | --- | --- | --- |
| $\Delta\mu_{Ca}$ | -4.00 | -6.15 | -4.02 | -6.16 |
| $\Delta\mu_{Cd}$ | 0.0 | -2.15 | 0.0 | -2.14 |
| $\Delta\mu_{O}$ | -2.15 | 0.0 | -2.14 | 0.0 |
| $\Delta\mu_{Na}$ | -0.98 | -2.28 | -0.94 | -2.28 |
| Competing phase | $Ca_3CdO_4$, Cd | $Ca_3CdO_4$ | $CaCd_3O_4$, Cd | $CaCd_3O_4$, $O_2(g)$ |
| Impurity phase | $Na_2CdO_2$ | $Na_2O_2$ | $Na_2CdO_2$ | $Na_2O_2$ |

Table S8: Considered competing phases for $CrCdO_4$.

|  | Competing phases |
| --- | --- |
| $CrCdO_4$ | $O_2(g)$, Cd, CdO, Cr, $Cr_2CdO_4$, $Cr_2O_3$, $CrCdO_4$, $CrO_2$ |
| Na | Na, $Na_{14}Cd_2O_9$, $Na_2CdO_2$, $Na_2Cr_2O_7$, $Na_2CrO_4$, $Na_2ONa_2O_2$, $Na_3Cd$, $NaCd_3$, $NaCr_3O_8$, $NaCrO_2$, $NaO_2$ |

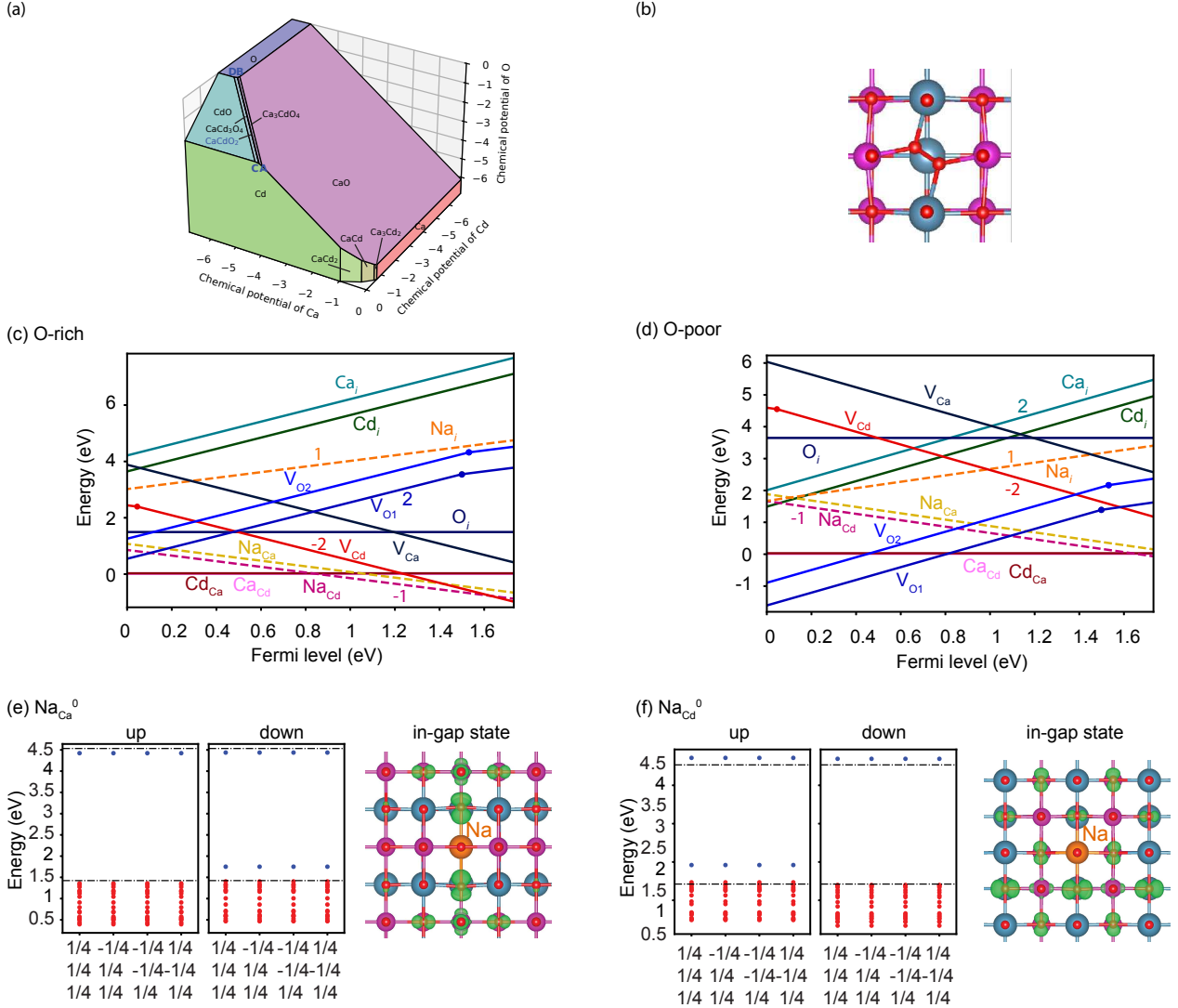

Figure S9: (a) Chemical potential diagram (CPD) of Ca-Cd-O ternary system. (b) Dimer structure for oxygen interstitial. (c-d) Formation energies of native defects and Na dopants in $L1_0$-$CaCdO_2$ as function of the Fermi level at oxygen-rich (B) and oxygen-poor (A) chemical potentials. (e-f) Single-particle levels at some $k$-points in the $L1_0$–$CaCdO_2$ and the corresponding of the squared wave functions for the in-gap states for the defect $Na_{Ca}^0$ and $Na_{Cd}^0$. Isosurfaces correspond to 5% of the maxima in each case.

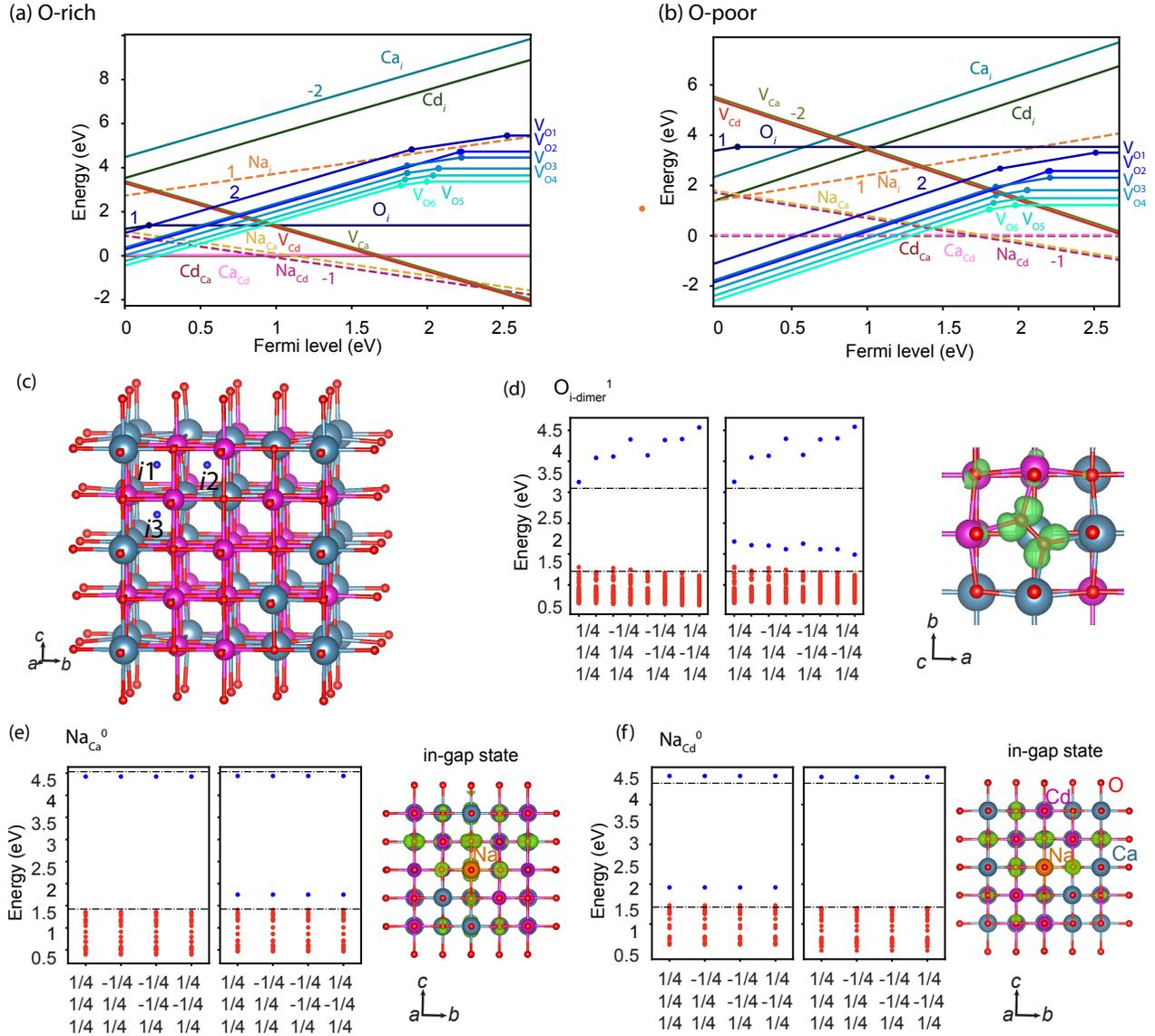

Figure S10: (a-b) Formation energies of native defects and Na dopants in SQS-CaCdO$_2$ model as function of the Fermi level at oxygen-rich (B) and oxygen-poor (A) chemical potentials. (c) Crystal structure with the interstitial sites are shown as a black spheres. (d) Single-particle levels at some $k$-points in the supercell models in dimer structure for oxygen interstitial defect with charge +1 and the corresponding of the squared wave functions for in-gap state. (e-f) Similar (d) but for Na$_{Ca}^0$ and Na$_{Cd}^0$.

13

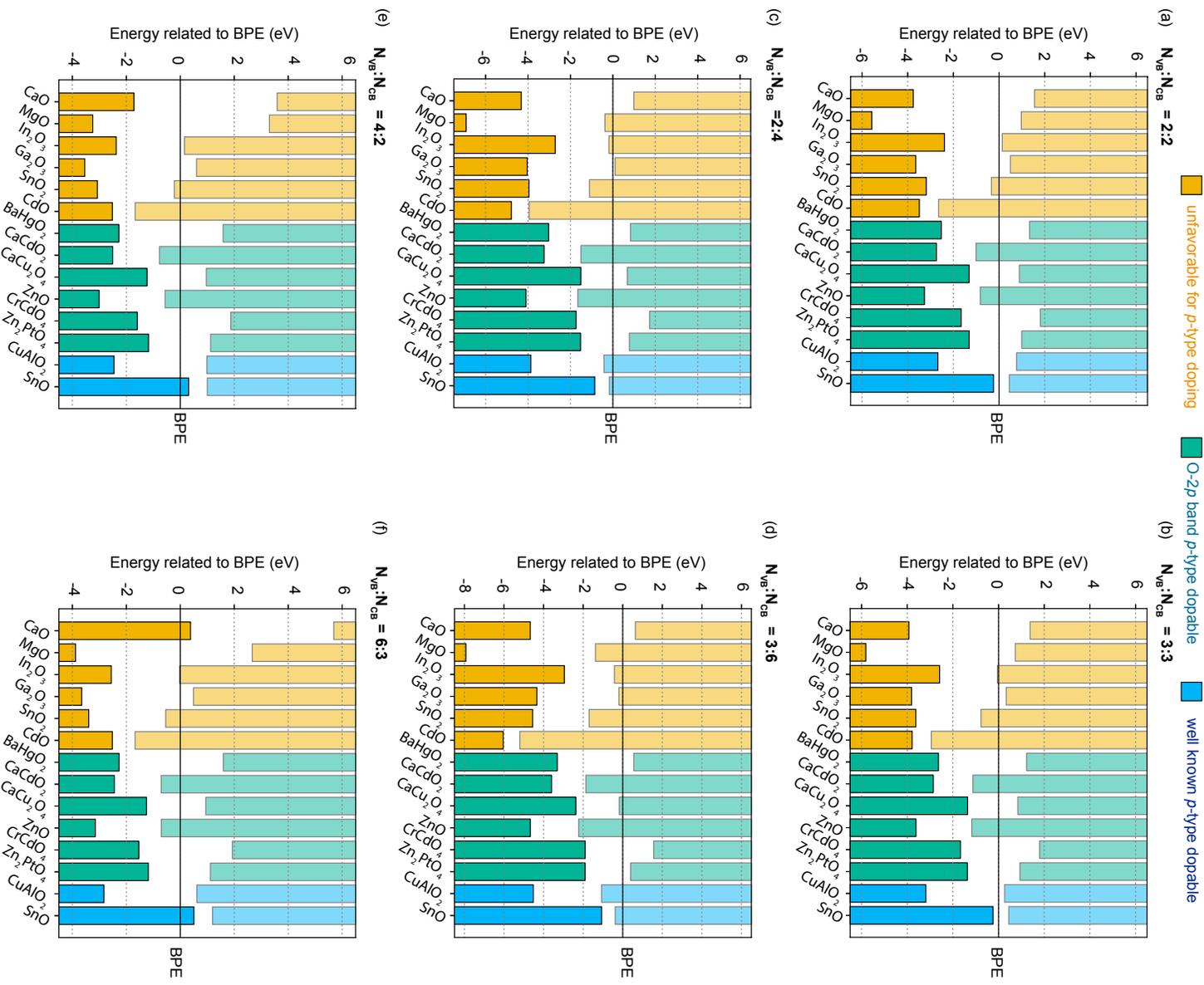

Figure S11: Band alignment obtained using HSE06, showing VBM across three group the same Fig. 6 in main text but aligned to branch point energy (BPE) with changing number of bands averaged over in valence band ($N_{VB}$) and in conduction band ($N_{CB}$).

14

Table S9: Considered competing phases for CaCdO$_2$.

| | Competing phases |
|---|---|
| CaCdO$_2$ | O$_2$(g), Ca, Ca$_3$Cd$_2$, Ca$_3$CdO$_4$, CaCd, CaCd$_2$, CaCd$_3$O$_4$, CaCdO$_2$, CaO, Cd, CdO |
| Na | Na, Na$_{14}$Cd$_2$O$_9$, Na$_2$CdO$_2$, Na$_2$O, Na$_2$O$_2$, Na$_3$Cd, NaCd$_3$, NaO$_2$ |

Table S10: Calculated sites for Na-substitution defects, cation vacancies, cation interstitials, and oxygen-dimer interstitials in the CaCdO$_2$ SQS model.

| Sites | Fractional coordinates |
|---|---|
| Cd1 | (1, 1/2, 1/2) |
| Cd2 | (0, 1/2, 0) |
| Cd3 | (1/2, 1/2, 1/2) |
| Ca1 | (1, 0, 1/2) |
| Ca2 | (1, 0, 1) |
| Ca3 | (1/2, 1, 1/2) |
| Na$_{i1}$,Ca$_{i1}$,Cd$_{i1}$ | (3/8, 1/8, 5/8) |
| Na$_{i2}$,Ca$_{i2}$,Cd$_{i2}$ | (3/8, 1/8, 7/8) |
| Na$_{i3}$,Ca$_{i3}$,Cd$_{i3}$ | (3/8, 3/8, 7/8) |
| O–dimer1 | (-0.05, 0.05, 0.255), (0.05, -0.05, 0.245) |
| O–dimer2 | (0.05, -0.05, 0.755), (-0.05, 0.05, 0.745) |
| O–dimer3 | (0.05, 0.55, 0.255), (0.05, 0.45, 0.245) |

Table S11: Calculated equilibrium Fermi level from the valence band maximum, defect, and carrier concentrations under the O-rich conditions at 1373 K in L1$_0$-CaCdO$_2$ are provided in units of cm$^{-3}$.

| $T$ (K) | $E_f$ (eV) | $p$ | $n$ | Ca$_{Cd}^0$ | Cd$_{Ca}^0$ | Na$_{Cd}^{-1}$ | Na$_{Ca}^{-1}$ | V$_{O1}^{2+}$ |
|---|---|---|---|---|---|---|---|---|
| 1373 | 0.268 | $1 \times 10^{20}$ | $5 \times 10^{13}$ | $2 \times 10^{22}$ | $2 \times 10^{22}$ | $1 \times 10^{20}$ | $2 \times 10^{19}$ | $2 \times 10^{18}$ |

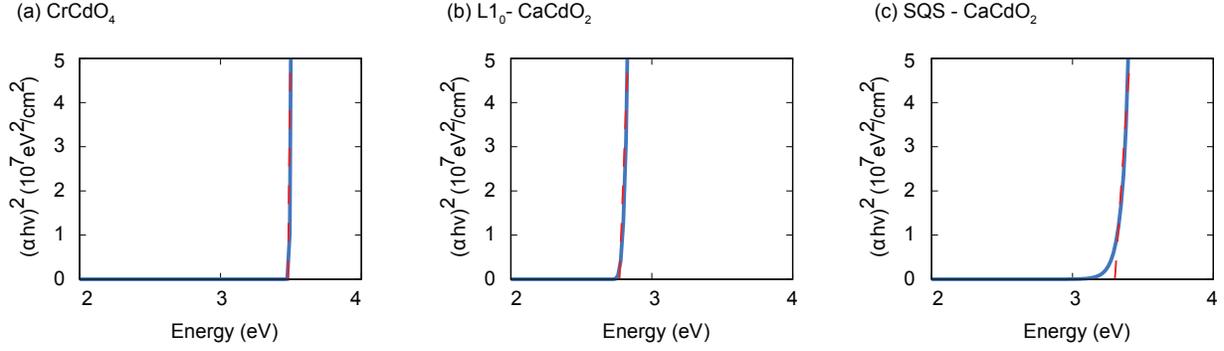

Figure S12: Tauc plots for the CrCdO$_4$, L1$_0$–CaCdO$_2$ and SQS–CaCdO$_2$.

electrical conduction in transparent thin films of CuAlO$_2$. *Nature* **1997**, *389*, 939–942.

(6) Varley, J. B.; Lordi, V.; Miglio, A.; Hautier, G. Electronic structure and defect properties of B$_6$O from hybrid functional and many-body perturbation theory calculations: A possible ambipolar transparent conductor. *Phys. Rev. B* **2014**, *90*, 045205.

(7) Akashi, T.; Itoh, T.; Gunjishima, I.; Masumoto, H.; Goto, T. Thermoelectric Properties of Hot-pressed Boron Suboxide (B$_6$O). *Materials Transactions* **2002**, *43*, 1719–1723.

(8) Kumagai, Y. Computational Screening of *p*-Type Transparent Conducting Oxides Using the Optical Absorption Spectra and Oxygen-Vacancy Formation Energies. *Phys. Rev. Appl.* **2023**, *19*, 034063.



\end{document}